\documentclass[a4paper,fleqn]{cas-dc}

\usepackage[numbers]{natbib}
\usepackage{subcaption}
\usepackage{float}
\usepackage{capt-of}

\def\tsc#1{\csdef{#1}{\textsc{\lowercase{#1}}\xspace}}
\tsc{WGM}
\tsc{QE}
\tsc{EP}
\tsc{PMS}
\tsc{BEC}
\tsc{DE}

\let\WriteBookmarks\relax

\begin{document}
\let\WriteBookmarks\relax
\def\floatpagepagefraction{1}
\def\textpagefraction{.001}
\shorttitle{Unified Steady-state Analysis}
\shortauthors{T. Gao et~al.}

\title [mode = title]{Unifying Power Flow and Electromagnetic Transient Modeling} 

\tnotetext[1]{This research did not receive any specific grant from funding agencies in the public, commercial, or not-for-profit sectors.}

\author[1]{Tina Gao}
\credit{Methodology, conceptualization, investigation, writing – original draft}
\cormark[1]
\author[1]{Aayushya Agarwal}
\credit{Conceptualization, investigation, writing --- review and editing}
\author[1]{Timothy McNamara}
\credit{conceptualization}
\author[1]{Lawrence Pileggi}
\credit{conceptualization, writing --- review and editing, funding acquisition}

\affiliation[1]{organization={Department of Electrical and Computer Engineering, Carnegie Mellon University}, 
                city={Pittsburgh},
                postcode={15213}, 
                state={Pennsylvania},
                country={United States}}

\cortext[cor1]{Corresponding author. Email address: tinagao@andrew.cmu.edu}

\begin{abstract}
Grid tools are separated by timescales: steady-state analysis is performed by power flow (PF), whereas the fastest dynamics are captured by electromagnetic transient (EMT) simulation. Although operating at varying timescales, different tools should produce consistent results when analyzing the same grid conditions. However, the PF steady-state solution does not match the time-to-infinity EMT response. This mismatch exists because simplified device models in PF are inconsistent with the ``ground truth" EMT models derived from first principles. As inverter-based resources (IBRs) and data centers strip away inertia and operating reserves, approximations used in PF risk grid security by missing violations. To address this, we introduce a steady-state framework that uses full-physics EMT models for high-fidelity steady-state grid analysis. The challenge lies in systematically formulating algebraic frequency-domain representations of dynamic power devices that are naturally described in state-space form. Our tool, SALT (Steady-state After Last Transients), constructs steady-state representations that exactly capture grid device physics when embedded into transmission networks. The approach exploits the structural properties of power device models and the single-harmonic, balanced nature of the transmission system. Results demonstrate SALT achieves EMT steady-state accuracy while delivering a $1\times10^6$ times speedup. We demonstrate that SALT can capture security threats in contingency scenarios without exception --- whereas PF-based analyses reported 75\% fewer rated line violations, 18\% fewer Q-limit violations, and 7\% fewer voltage violations.
\end{abstract}

\begin{graphicalabstract}
\includegraphics[width=\linewidth]{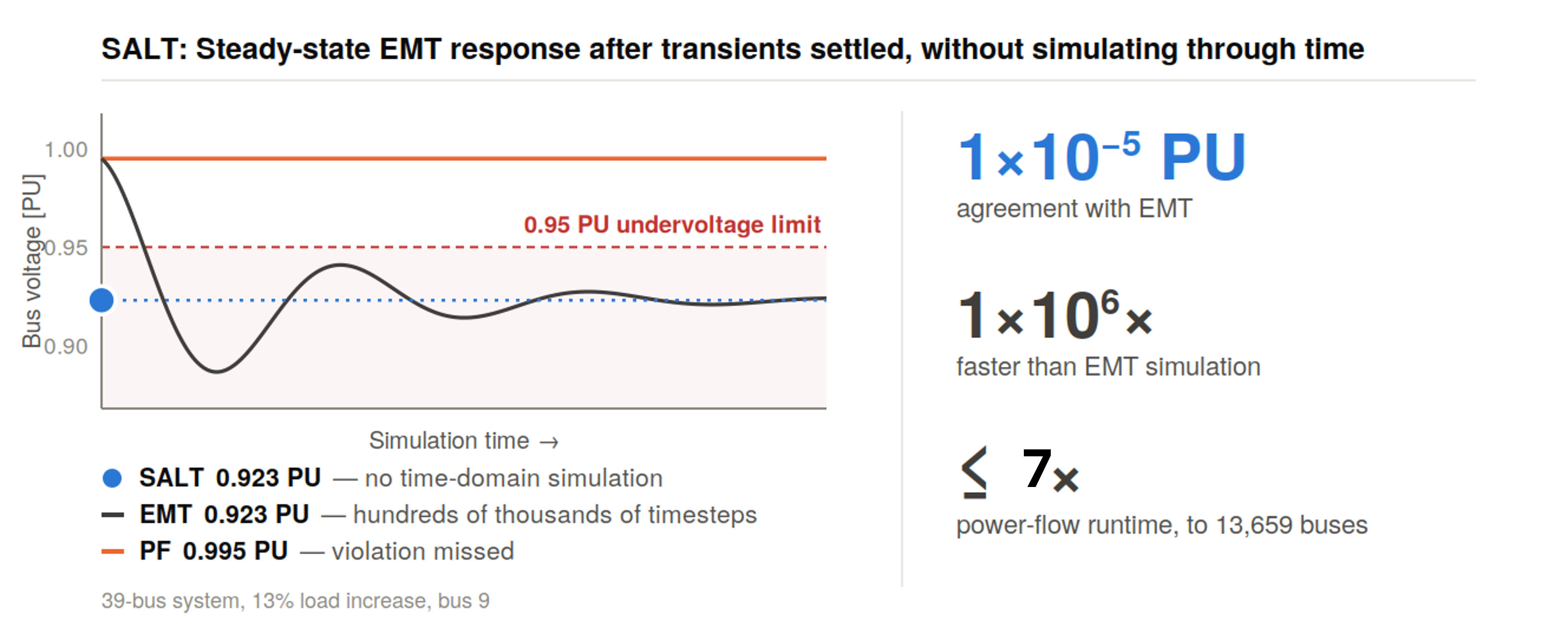}
\end{graphicalabstract}

\begin{highlights}
\item SALT is a fully physics-based steady-state solver for transmission analysis
\item Achieves electromagnetic transient (EMT) fidelity at computational efficiency of power flow (PF)
\item  Captures all security risks whereas PF reported 75\% fewer line violations in contingency analysis

\end{highlights}


\maketitle

\section{Introduction}

Historically, power systems analysis has relied on different algorithms to study phenomena of different timescales. For steady-state analysis, power flow (PF) assesses static power balance using abstractions in the frequency domain. For capturing fast microsecond dynamics, electromagnetic transient (EMT) simulation computes the point-on-wave response using time domain models derived from physical principles and is considered the “ground-truth” simulator. Transient stability analysis (TSA) occupies the middle, assessing grid synchronism through time-varying phasors around 60 Hz.

Although operating at varying timescales, different tools modeling the same grid conditions should compute consistent results. However, as depicted in Figure~\ref{figMismatchSoln}, the steady-state solution calculated by PF does not match the settled response of the EMT simulation. While EMT identifies the post-contingency voltage settling outside the nominal region, PF fails to detect this violation, risking
equipment damage and system instability.

The inconsistency arises from approximations in PF models, initially developed for constraints of early computers \cite{PF,PF_constraints}. Traditional PF abstracts all devices to static constant power-voltage (PV) generators or constant real-reactive-power (PQ) loads, and a slack generator artificially absorbs system power mismatches, sustaining the idealized operating frequency of 60 Hz. Conversely, EMT device models are designed to capture wave phenomenon, resulting in dynamics and nonlinearities derived from first principles \cite{EMT}. The system frequency deviates as a response to mismatches between production and demand.

\par\addvspace{\intextsep}%
{\centering\captionsetup{type=figure}%
\centering
\includegraphics[width=0.7\linewidth]{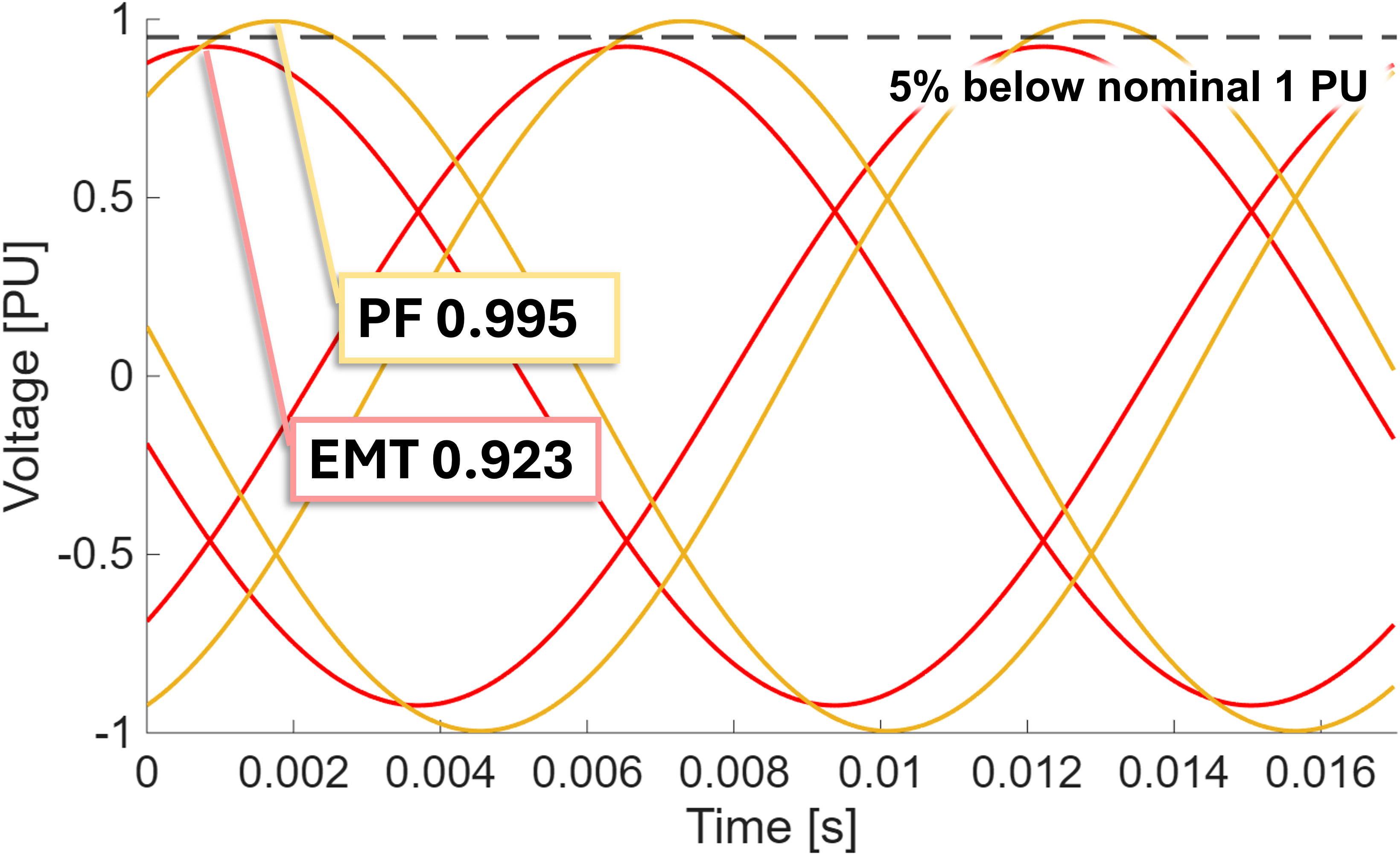}
\par\caption{\textit{\textbf{PF computes steady-state solution disparate from corresponding EMT simulation-to-infinity, failing to detect undervoltage violation.} 39-bus system \cite{ParaEMT1,ParaEMT2} (all generating units are inverter-based resources except at bus 30, 31, and 39) incurs 13\% load increase. At bus 9, PF computes a voltage magnitude ($0.995$ PU) within operating range (5\% of 1 PU), failing to catch the undervoltage detected by EMT (0.923 PU).}}
\label{figMismatchSoln}
\par}
\addvspace{\intextsep}\par

As the grid grows fragile due to reduced inertia from the penetration of inverter-based resources (IBRs) and shrinking reserve margins from data center growth \cite{gridInertia1, gridInertia2,dataCenter}, the disparity between PF and EMT poses a security risk for steady-state analysis. The modern grid is less forgiving of approximations, as errors once absorbed by inertia and reserve margins can now cascade into security concerns --- this is attested by the fact that stability analyses traditionally performed by TSA are increasingly replaced by EMT \cite{gridInertiaEMT}. Additionally, PF solutions are used to initialize EMT simulations, but inconsistencies between PF and EMT models require extra computation to resolve errors injected into initialization \cite{EMTinit}. This motivates the need for a steady-state analysis that is consistent with EMT solutions.

Currently, computing the EMT steady-state solution is inefficient. Unlike frequency-domain PF-type solvers that compute the grid at a single state, EMT is a time-domain simulator and evaluates state-space equations at discrete timesteps until transients decay to negligible quantities. Due to the stiffness of grid differential equations, microsecond stepsizes are required to resolve the dynamics of the fastest devices, while decay to steady-state is limited by slowest device time constants of minutes to hours \cite{Cosim1}. Settling can be further prolonged by controllers fighting. Harmonic Balance (HB) methods \cite{HB,HBPower1,HBPower2,HBPower3} have been proposed as a general method to map nonlinear device models to the frequency-domain. However, the memory of dynamic models requires HB to evaluate the transient response of each device to steady-state, inheriting the burden of EMT simulations. This motivates the need for a steady-state analysis with EMT fidelity that retains the efficiency of PF.

We present SALT (Steady-state After Last Transients) as an alternative to PF that systematically constructs steady-state models that are unified with physics-based EMT device models for transmission network analysis. SALT converts EMT differential equations into algebraic steady-state forms by exploiting the structure of power devices \cite{EmtDq0Struct} for single-harmonic, balanced systems \cite{Kundur}. Unlike HB and transient methods, SALT models are solved fully in the frequency-domain, eliminating the burden of stepping through time.

Our experiments show that SALT solutions are within local truncation error (LTE) of settled EMT waveforms while delivering around a $1\times10^6$ times speedup and maintaining the computational complexity of PF. SALT captured all security risks, while PF-based methods reported 75\% fewer rated line violations, as well as less Q-limit and voltage violations, in a contingency analysis for a 39-bus network.

\section{Literature Review}

Despite being the backbone of grid operation and planning, accurate steady-state analysis with computational efficiency is difficult because, in theory, it involves taking the EMT response to infinite time. Existing research can be grouped in three approaches: enhancing assumptions in PF, improving the speed of EMT simulation, and bridging steady-state frameworks with EMT through co-simulation.

\subsection{PF Enhancements}

Previous works improved PF assumptions by defining all 3-phases \cite{PFunbalanced} and modeling frequency deviations via distributed slack, capturing power share among generation units \cite{PFfreqDev1} and modeling frequency control responses dynamically \cite{PFfreqDevAgarwald}. Other post-contingency controls have been included to model the re-allocation of power \cite{PFcontrol1} and response to trips \cite{PFcontrol2}. Static ZIP loads are fitted with voltage dependency \cite{loadSummary}. However, although existing research has expanded PF to accommodate more physical phenomenon, the device models remain abstract. Conversely, SALT utilizes exact physics-based steady-state models.

\subsection{EMT Speedup}

To accelerate EMT simulations, researchers have mitigated timescale stiffness by co-simulating slower dynamics with TSA \cite{Cosim1,Cosim2,Cosim3} and expedited individual matrix solves by decoupling the system, allowing for parallelization \cite{EMTSS,ParaEMT1,ParaEMT2}. However, this inherently still bears the computational burden of simulating through time. Unlike frequency-domain solvers that use Newton-Raphson (NR) to iterate on steady-state snapshots of the grid, EMT and other transient simulators advance the system until transients decay, magnifying the computational burden as each NR inner-loop is solved via numerical integration for every timepoint in the outer-loop. TSA uses larger timesteps than EMT but inherently incurs the computational burden of transient simulation while losing fidelity \cite{TSA}.

\subsection{Harmonic Balance: Frequency-domain and EMT Co-simulation}

Previous efforts to unify steady-state and transient grid analysis use the equivalent circuit framework \cite{SugarUnification,EqCircuit1,EqCircuit2,EqCircuit3} to apply HB, bridging time-domain models and the frequency-domain steady-state problem through the Discrete Fourier Transform (DFT) \cite{HB}. However, the DFT introduces the computational burden of conversion matrices which, although necessary for multi-harmonic systems, adds arithmetic overhead when evaluating transmission networks whose higher frequencies are attenuated from loss \cite{Kundur}. More critically, salient grid components such as synchronous generators (SGs), IBRs, and induction motors (IMs) are inherently dynamic. In existing literature, this memoryless limitation of HB is omitted \cite{HBPower1,HBPower2} or managed by computing the DFT on top of a settled time-domain simulation for each non-memoryless device \cite{HBPower3}, reintroducing the cost of EMT simulation.

\section{Unified Analysis}

Conversely, SALT bypasses the aforementioned limitations by exploiting structural properties of EMT devices in transmission networks and avoiding any time-domain analyses. SALT constructs a steady-state framework through a frequency-domain analysis of EMT state-space equations, ensuring that the steady-state operating point is consistent with grid physics. The system is characterized by the state-space equations:
\begin{equation}
    \dot{\textbf{q}}(\textbf{v}_t) + \textbf{Rv}_t + \textbf{i}(\textbf{v}_t) = \textbf{0}
\label{eq:HB_TD}
\end{equation}

\noindent where $\mathbf{v}_t$ is the vector of 3-phase voltages at time $t$, $\mathbf{q}(\mathbf{v}_t)$ is a vector of electric charges, $\textbf{R}$ is a matrix of inverse resistances, and $\mathbf{i}(\mathbf{v}_t)$ is the vector of current source injections from power device models derived from first-principles.

The steady-state behavior of the grid is defined as the settled response from a simulation-to-infinity, naturally represented in the frequency-domain because transients have decayed into periodic waveforms. The frequency-domain response is computed by:
\begin{align}
\tilde{\mathbf{V}} &= \mathcal{F}\{ \mathbf{v_t} \}, \quad \quad t \rightarrow \infty
\label{unifyDef}
\end{align}
\noindent where $\tilde{\mathbf{V}}$ are complex bus voltages and $\mathcal{F}$ is the Fourier transform. The frequency-domain result of \eqref{eq:HB_TD} is:
\begin{equation}
    \Omega \textbf{Q}(\tilde{\mathbf{V}}, \omega) + \textbf{Y}(\omega)\tilde{\mathbf{V}} + \textbf{I}(\tilde{\mathbf{V}}, \omega) = \textbf{0}
\label{eq:HB_FD}
\end{equation}

\noindent where  $\Omega=j\omega$ is the frequency-domain derivative, $\mathbf{Q}(\tilde{\mathbf{V}}, \omega)$ is the frequency-domain counterpart for charges, $\mathbf{Y}(\omega)$ is the admittance, $\mathbf{I}(\tilde{\mathbf{V}}, \omega)$ is a frequency-domain model of nonlinear power devices, and $\omega$ holds Fourier frequencies. In transmission systems where non-fundamental frequencies decay aggressively, the spectrum of frequencies is reduced to a single system frequency, $\omega_{s0}$, which is included as a state variable to model frequency deviation. A reference bus defining the phase is added for the full-rank of the system:
\begin{align}
V_{a,ref,im} = V_{a,ref,re}\tan(\phi_{a,slack})
\label{refBus}
\end{align}
\noindent where $V_{a,ref,re/im}$ is the complex phase-A voltage of the reference bus and $\phi_{a,slack}$ is the reference angle.

SALT builds frequency-domain representations of devices on a component-by-component basis. Linear steady-state models are captured trivially by $\textbf{Y}(\omega_{s0})\tilde{\mathbf{V}}$, where constant multipliers remain unchanged while the derivative is replaced by $j\omega_{s0}$. In theory, nonlinear terms $\Omega \textbf{Q}(\tilde{\mathbf{V}}, \omega_{s0})+\textbf{I}(\tilde{\mathbf{V}}, \omega_{s_0})$ are solved by analytically evaluating $\mathcal{F}\{\dot{\textbf{q}}(\textbf{v}_t)+ \textbf{i}(\textbf{v}_t) \}$, but this generally admits no closed-form solution due to dynamic, nonlinear couplings. HB iterates between a settled response of $\dot{\textbf{q}}(\textbf{v}_t)+ \textbf{i}(\textbf{v}_t)$ and the frequency-domain via the DFT, but this carries overhead from conversion matrices and time-domain simulation. The next Section-4 describes building SALT models to bypass these limitations.

\section{SALT Nonlinear Modeling}

SALT models are constructed by exploiting the $dq$-structure shared by EMT devices under balanced single-harmonic conditions, motivating the following definitions which provide the foundation of SALT.

\subsection{Definitions}

\textit{Definition 1 ($dq$-structure): EMT grid device models are characterized by the $dq$-structure in Figure~\ref{figureEmtStruct}, which consists of a $dq$-transform interfacing between 3-phase network quantities and $dq$-quantities driving the internal $dq$-circuit.}

The $dq$-transform projects sinusoidal grid quantities onto a synchronous $dq$-reference frame. Used to conveniently render inductances constant in the $dq$-circuit, the $dq$-transform is a sufficiently universal feature of EMT models \cite{EmtDq0Struct} for SALT to justifiably use in constructing steady-state models.

\par\addvspace{\intextsep}%
{\centering\captionsetup{type=figure}%
\centering
\vspace{-0.2cm}
\includegraphics[width=1\linewidth]{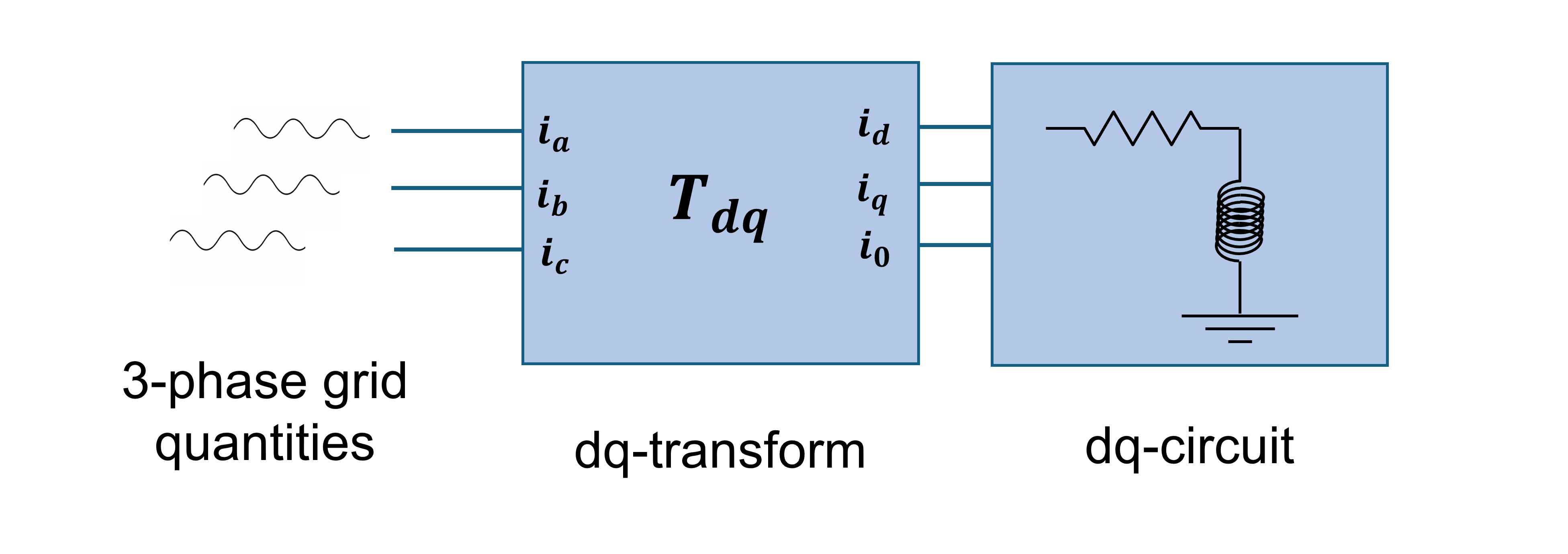}
\vspace{-0.75cm}
\par\caption{\textit{\textbf{\textit{dq}-structure characterizing EMT device models.}}}
\label{figureEmtStruct}
\par}
\addvspace{\intextsep}\par

Mathematically, the $dq$-transform and its inverse are characterized by \cite{Kundur}:

\begin{subequations}
\begin{align}
    \begin{bmatrix} i_{d,t} \\ i_{q,t} \\ i_{0,t} \end{bmatrix}
    &=
    \frac{2}{3}
    \scalebox{0.75}{$\displaystyle
    \begin{bmatrix}
        \cos\theta_t & \cos\left(\theta_t - \frac{2\pi}{3}\right) & \cos\left(\theta_t + \frac{2\pi}{3}\right) \\[1ex]
        -\sin\theta_t & -\sin\left(\theta_t - \frac{2\pi}{3}\right) & -\sin\left(\theta_t + \frac{2\pi}{3}\right) \\[1ex]
        \frac{1}{2} & \frac{1}{2} & \frac{1}{2}
    \end{bmatrix}
    $}
    \begin{bmatrix} i_{a,t} \\ i_{b,t} \\ i_{c,t} \end{bmatrix}
    \label{eq:park} \\[2ex]
    \begin{bmatrix} i_{a,t} \\ i_{b,t} \\ i_{c,t} \end{bmatrix}
    &=
    \scalebox{0.85}{$\displaystyle
    \begin{bmatrix}
        \cos\theta_t & -\sin\theta_t & 1 \\[1ex]
        \cos\left(\theta_t - \frac{2\pi}{3}\right) & -\sin\left(\theta_t - \frac{2\pi}{3}\right) & 1 \\[1ex]
        \cos\left(\theta_t + \frac{2\pi}{3}\right) & -\sin\left(\theta_t + \frac{2\pi}{3}\right) & 1
    \end{bmatrix}
    $}
    \begin{bmatrix} i_{d,t} \\ i_{q,t} \\ i_{0,t} \end{bmatrix}
    \label{eq:invPark}
\end{align}
\label{eq:invPark_all}
\end{subequations}

\noindent where $i_{a/b/c,t}$ are 3-phase grid currents, $i_{d/q/0,t}$ are their projections onto the $dq$-frame. For SGs, $\theta_t$ is the angle by which the $d$-axis leads the phase-A winding, and the $d$-axis is aligned with the magnetic axis of the rotor and lags the $q$-axis by 90$^{\circ}$. Alternative transform formulations for different devices (e.g. quantity invariance, axis alignment and positioning) \cite{Krause,EmtDq0Struct,EmtDq0} map to \eqref{eq:invPark_all} via trivial linear transformations of the $dq$-quantities such as scaling, row swapping, and sign changes. With slight abuse in notation, $i_{d/q/0,t}$ is treated as the transformed quantities and $\theta_t$ as a general reference, allowing \eqref{eq:invPark_all} to be used for all devices without loss of generality.

The $dq$-circuit in Figure~\ref{figureEmtStruct}, whose inputs $i_{d/q,t}$ are projections computed by the $dq$-transform, in its general form is a nonlinear set of differential equations:
\begin{equation}
\resizebox{0.9\linewidth}{!}{$%
    \begin{bmatrix} 
    \mathbf{C}_{DQ} & \\ 
    & \mathbf{L}_{DQ} 
    \end{bmatrix} 
    \dot{\begin{bmatrix} 
    \mathbf{v}_{DQ,t} \\ 
    \mathbf{i}_{DQ,t} 
    \end{bmatrix}}
    + \begin{bmatrix} 
    \mathbf{G}_{DQ} & \\ 
    & \mathbf{R}_{DQ} 
    \end{bmatrix} 
    \begin{bmatrix} 
    \mathbf{v}_{DQ,t} \\ 
    \mathbf{i}_{DQ,t} 
    \end{bmatrix}
    + \mathbf{F}_{DQ} \!\left( \begin{bmatrix} \mathbf{v}_{DQ,t} \\ \mathbf{i}_{DQ,t} \end{bmatrix} \right) 
    + \mathbf{E} 
    \begin{bmatrix} 
    i_{d,t} \\ 
    i_{q,t} \\ 
    i_{0,t} 
    \end{bmatrix} = \mathbf{0}
$}
\label{eq:id_internalCircuit}
\end{equation}

\noindent where $\mathbf{C}_{DQ}$ and $\mathbf{L}_{DQ}$ are capacitances and inductances for filters and windings, $\mathbf{G}_{DQ}$ and $\mathbf{R}_{DQ}$ are conductive and resistive losses, $\mathbf{F}_{DQ}$ is a function for internal nonlinearities, and $\mathbf{E}_{dq0}$ is a selection matrix for $dq$-inputs. Without loss of accuracy, SALT simplifies this general formulation by applying steady-state characteristics for transmission networks.

\textit{Definition 2 (transmission steady-state): The steady-state transmission network is single-harmonic and balanced.} 

This definition of steady-state is well-justified for transmission systems, where harmonics attenuate across long lines and balance is enforced to minimize loss\cite{Kundur}. 

Under these conditions, the time-domain 3-phase grid quantities $i_{a/b/c,t}$ are sinusoids with equal magnitude and frequency, with phases $120^{\circ}$ apart:
\begin{subequations}
\label{eq:SS_TD} 
\begin{align}
    i_{a,t} &= I_mcos(\omega_{s0}t+\phi_0) \\
    i_{b,t} &= I_mcos(\omega_{s0}t+\phi_0-\frac{2\pi}{3}) \\
    i_{c,t} &= I_mcos(\omega_{s0}t+\phi_0+\frac{2\pi}{3})
\end{align}
\end{subequations}
\noindent where $I_m$ is the amplitude, $\omega_{s0}$ is the frequency, and $\phi_0$ is the phase. The corresponding frequency domain representation is:
\begin{subequations}
\label{eq:SS_FD}
\begin{align}
    \tilde{I_a} &= I_me^{j\phi_0} \\
    \tilde{I}_b &= \tilde{I}_ae^{-j\frac{2\pi}{3}} \\
    \tilde{I}_c &= \tilde{I}_ae^{j\frac{2\pi}{3}}
\end{align}
\end{subequations}

\noindent Under these single-harmonic and balanced conditions, EMT device reference angles increase at synchronous speed:
\begin{equation}
\label{eq:theta_synch_speed}
\theta_t = \omega_{s0}t + \theta_0
\end{equation}

\subsection{Methodology}

By integrating classical steady-state results of $dq$-based machines \cite{Kundur} with HB theory, SALT contributes a generalized framework for modeling EMT devices in the frequency-domain.

\textit{First, recognize that the $dq$-currents entering the $dq$-circuit as constants in steady-state.}

Physically, this occurs in steady-state because the reference frame attached to the rotor spins at the same synchronous speed as grid quantities, locking their relative angle. This is equivalently derived by substituting the steady-state currents \eqref{eq:SS_TD} and reference angle constraint \eqref{eq:theta_synch_speed} into the $dq$-transform \eqref{eq:park} \cite{Kundur}. Frequency terms cancel out, rendering $i_{d/q,t}$ constant and evaluating $i_{0,t}(t)$ to zero:
\begin{equation}
    i_{d,t} = i_{d} \quad 
    i_{q,t} = i_{q} \quad 
    i_{0,t} = 0 \quad 
    \label{eq:step_1_conclusions}
\end{equation}

\textit{Second, as a result of constant $dq$-inputs, re-formulate the dynamic $dq$-circuit as a memoryless system}.

When the inputs to a circuit are direct current (DC), capacitors are opened and inductors are shorted. As the transients represented by the first term of \eqref{eq:id_internalCircuit} are driven to zero, the internal states of the $dq$-circuit, $\mathbf{v}_{DQ}$ and $\mathbf{i}_{DQ}$, can be effectively solved as a memoryless system:
\begin{equation}
\resizebox{0.7\linewidth}{!}{$
    \begin{bmatrix} 
    \mathbf{G}_{DQ} & \\ 
    & \mathbf{R}_{DQ} 
    \end{bmatrix} 
    \begin{bmatrix} 
    \mathbf{v}_{DQ} \\ 
    \mathbf{i}_{DQ} 
    \end{bmatrix}
    + \mathbf{F}_{DQ} \!\left( \begin{bmatrix} \mathbf{v}_{DQ} \\ \mathbf{i}_{DQ} \end{bmatrix} \right) 
    + \mathbf{E}_{dq} 
    \begin{bmatrix} 
    i_{d} \\ 
    i_{q}
    \end{bmatrix} = \mathbf{0}
$}
\label{eq:step_2_conclusions}
\end{equation}

At this point, the time-domain EMT model has been re-characterized as two algebraic subcircuits: an internal $dq$-circuit that is memoryless due to being driven by DC inputs, and a $dq$-transform subcircuit that interfaces with the grid. Because the trigonometric terms of the $dq$-transform make the frequency response non-trivial to derive, conventional HB methods would now naively compute the device terminal waveforms $i_{a/b/c,t}$ in the frequency-domain via the DFT, incurring the computational burden of evaluating dense conversion matrices coupled with the highly nonlinear $dq$-transform. Conversely, SALT circumvents the $dq$-transform and DFT with a simple phase shift. This final step in SALT, given in context with the previous steps, is illustrated in Figure \ref{figureSALTStructure}.

\textit{Finally, embed the memoryless $dq$-circuit equations from step two into steady-state analysis by bringing its input signals $i_{d/q}$ into the frequency-domain via a phase shift.}

The lack of needing an explicit time-to-frequency-domain transformation of the $dq$-circuit, except for a phase shift on $i_{d/q}$, comes from recognizing that the $dq$-quantities are inherently a variant of frequency-domain phasors. The $dq$-transform is functionally analogous to a frequency-domain transform --- the $dq$-transform and DFT both project time-domain signals onto a different basis, and in steady-state, the $dq$-transform basis is also composed of periodic sinusoids due to the reference frame rotating at constant speed \cite{Kundur}.

\par\addvspace{\intextsep}%
{\centering\captionsetup{type=figure}%
\centering
\vspace{-0.25cm}
\includegraphics[width=0.85\linewidth]{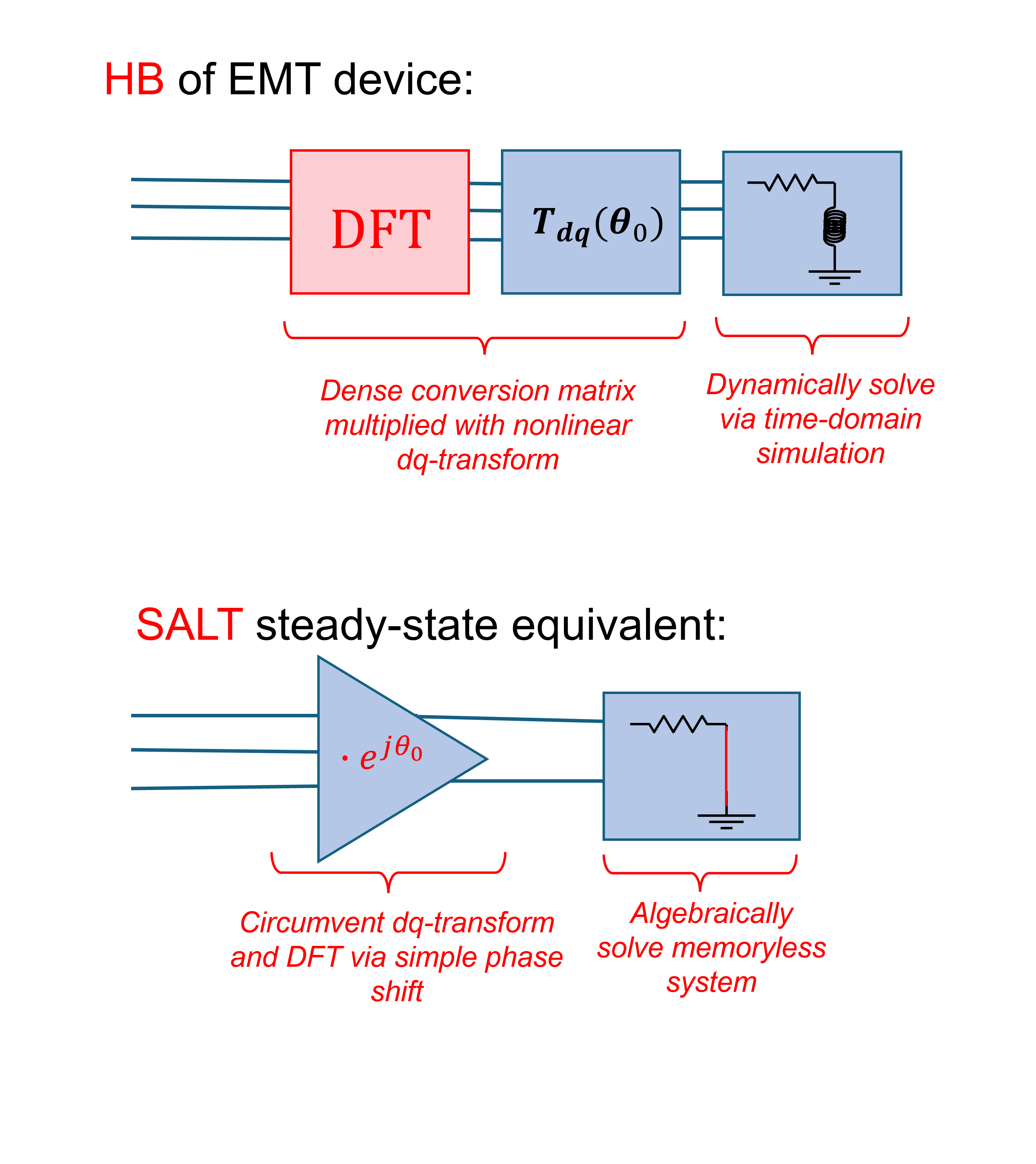}
\vspace{-1.1cm}
\par\caption{\textit{\textbf{SALT analytically connects the memoryless $dq$-circuit with the frequency-domain via phase shift.} The DFT and time-domain simulation are circumvented.}}
\label{figureSALTStructure}
\par}
\addvspace{\intextsep}\par

SALT formally reaches the same conclusion by bringing the $dq$-structures into the frequency-domain through an analytical evaluation of the Fourier transform on the terminal currents $i_{a/b/c,t}$. 

\textit{Claim ($dq$-quantities with phase shift): Given the steady-state assumptions for transmission system, the frequency-domain grid quantities are equal to the time-domain $dq$-quantities with a phase shift}.

\textit{Proof of claim:} The proof is specific to the phase-A grid quantity without loss of generality. Based on the balanced assumption, phases-B and -C can be proved identically by first converting them to phase-A using phase shifts derived from the balanced assumption \eqref{eq:SS_FD}.

First, calculate the time-domain phase-A current. Evaluate the inverse Park transform \eqref{eq:invPark} after substituting in the constant $i_{d/q/0}$ inputs derived in the first step \eqref{eq:step_1_conclusions} and constraining $\theta_t$ to a steady-state response given by \eqref{eq:theta_synch_speed}, getting: 
\begin{equation*}
\begin{aligned}
\scalebox{1}{$\displaystyle
i_{a,t} = i_d \cos(\omega_{s0}t + \theta_0) - i_q \sin(\omega_{s0}t + \theta_0)
$}
\end{aligned}
\end{equation*}

\noindent When sampled with $S$ points per period, $i_{a,t}$ becomes:
\begin{equation*}
\begin{aligned}
\scalebox{0.6}{$\displaystyle
i_{a,s} = i_d \cos\left(\frac{2\pi s}{S} + \theta_0\right) - i_q \sin\left(\frac{2\pi s}{S} + \theta_0\right)
$}
\end{aligned}
\end{equation*}

\noindent Expanding the trigonometric terms produces the following:
\begin{equation*}
\begin{aligned}
\scalebox{0.6}{$\displaystyle
i_{a,s} = i_d \left[ \cos\left(\frac{2\pi s}{S}\right)\cos(\theta_0) - \sin\left(\frac{2\pi s}{S}\right)\sin(\theta_0) \right]
- i_q \left[ \sin\left(\frac{2\pi s}{S}\right)\cos(\theta_0) + \cos\left(\frac{2\pi s}{S}\right)\sin(\theta_0) \right]
$}
\end{aligned}
\end{equation*}

\noindent Rearranging terms by the time-varying components (sinusoids with $\frac{2\pi s}{S}$) yields the sampled current as:
\begin{equation}
\begin{aligned}
\scalebox{0.65}{$\displaystyle
i_{a,s} = \left[ i_d\cos(\theta_0) - i_q\sin(\theta_0) \right] \cos\left(\frac{2\pi s}{S}\right) - \left[ i_d\sin(\theta_0) + i_q\cos(\theta_0) \right] \sin\left(\frac{2\pi s}{S}\right)
\label{eq:i_a_final}
$}
\end{aligned}
\end{equation}

Next, analytically evaluate the Fourier transform of $i_{a,s}$ \eqref{eq:i_a_final}. Here, the DFT is used instead of the continuous-time Fourier transform because the DFT is exact when no harmonics are truncated, but its discrete evaluation avoids scaling factors of $\pi$ associated with continuous integration limits. The DFT of a discrete signal $x(s)$ is defined as \cite{HBPower1}:
\begin{equation}
\begin{aligned}
X(k) = \begin{bmatrix} X_{re}(k) \\ X_{im}(k) \end{bmatrix} = \frac{2 - \delta(k)}{S} \sum_{s=0}^{S-1} \begin{bmatrix} \cos\left(\frac{2\pi k s}{S}\right) \\ -\sin\left(\frac{2\pi k s}{S}\right) \end{bmatrix} x(s)
\label{eq:DFT}
\end{aligned}
\end{equation}
\noindent where the window length is $S = 2K - 1$, and $K$ is the number of captured frequencies. For the single-harmonic case in SALT, the DFT is exact when $K=2$, capturing the fundamental system frequency $\omega_{s0}$ alongside a DC component. This results in a window of $S=3$ samples.

For the DC harmonic ($k=0$), the DFT \eqref{eq:DFT} of $i_{a,s}$ is calculated as such:
\begin{equation*}
\begin{aligned}
\scalebox{0.55}{$\displaystyle X_{i_a}(k=0) $} &= \scalebox{0.75}{$\displaystyle \frac{2 - \delta(0)}{S} \sum_{s=0}^{S-1} \begin{bmatrix} \cos(0) \\ -\sin(0) \end{bmatrix} i_a(s) $}
\end{aligned}
\end{equation*}

\noindent The scaling factor evaluates to $\frac{2 - \delta(0)}{S}=\frac{1}{S}$. Now, compute $\sin(0)=0$ and $\cos(0)=1$, substitute the full representation of $i_{a,t}$ \eqref{eq:i_a_final}, and compute again to get:

\begin{equation*}
\begin{aligned}
\scalebox{0.75}{$\displaystyle X_{i_a}(k=0) $} &= \scalebox{0.75}{$\displaystyle \frac{1}{S} \begin{bmatrix} \sum_{s=0}^{S-1} i_a(s) \\ 0 \end{bmatrix} $} \\
&= \scalebox{0.75}{$\displaystyle \frac{1}{S}\begin{bmatrix} \sum_{s=0}^{S-1} \left[ i_d\cos(\theta_0) - i_q\sin(\theta_0) \right] \cos(0) - [...] \sin(0) \\ 0 \end{bmatrix} $} \\
&= \scalebox{0.75}{$\displaystyle \frac{1}{S}\begin{bmatrix} \sum_{s=0}^{S-1} i_d\cos(\theta_0) - i_q\sin(\theta_0) \\ 0 \end{bmatrix} $}
\end{aligned}
\end{equation*}

\noindent Calculate the remaining terms as a summation of sinusoids over a full period, which evaluates the DC component as 0:

\begin{equation}
\begin{aligned}
\scalebox{0.65}{$
X_{i_a}(k=0)
=
\frac{1}{S}\begin{bmatrix} 
i_d\sum_{s=0}^{S-1} \cos(\theta_0) - i_q\sum_{s=0}^{S-1}\sin(\theta_0) \\ 0 \end{bmatrix}
= 
\frac{1}{S}\begin{bmatrix} 
i_d\cdot0 - i_q\cdot0 \\ 0 \end{bmatrix} = 
\begin{bmatrix} 0 \\ 0 \end{bmatrix}
$}
\label{eq:X0_Coeff}
\end{aligned}
\end{equation}

For the single harmonic ($k=1$), the DFT \eqref{eq:DFT} of $i_{a,s}$ is calculated as:
\begin{equation*}
\begin{aligned}
\scalebox{0.55}{$\displaystyle X_{i_a}(k=1) $} &= \scalebox{0.75}{$\displaystyle \frac{2 - \delta(1)}{S} \sum_{s=0}^{S-1} \begin{bmatrix} \cos\left(\frac{2\pi s}{S}\right) \\ -\sin\left(\frac{2\pi s}{S}\right) \end{bmatrix} i_{a,s} $}
\end{aligned}
\end{equation*}

\noindent The scaling factor evaluates to $\frac{2 - \delta(1)}{S}=\frac{2}{S}$. Substitute the full representation of $i_{a,t}$ \eqref{eq:i_a_final} and rearrange the summation:
\begin{equation*}
\scalebox{0.6}{$\displaystyle
\begin{aligned}
& X_{i_a}(k=1) = \\
& \frac{2}{S} \sum_{s=0}^{S-1} \begin{bmatrix}
\cos\left(\frac{2\pi s}{S}\right) \left( \left[ i_d\cos(\theta_0) - i_q\sin(\theta_0) \right]\cos\left(\frac{2\pi s}{S}\right) - \left[ i_d\sin(\theta_0) + i_q\cos(\theta_0) \right]\sin\left(\frac{2\pi s}{S}\right) \right) \\[2ex]
-\sin\left(\frac{2\pi s}{S}\right) \left( \left[ i_d\cos(\theta_0) - i_q\sin(\theta_0) \right]\cos\left(\frac{2\pi s}{S}\right) - \left[ i_d\sin(\theta_0) + i_q\cos(\theta_0) \right]\sin\left(\frac{2\pi s}{S}\right) \right)
\end{bmatrix} = \\[3ex]
&\frac{2}{S} \begin{bmatrix}
\left[ i_d\cos(\theta_0) - i_q\sin(\theta_0) \right] \sum_{s=0}^{S-1} \cos^2\left(\frac{2\pi s}{S}\right) - \left[ i_d\sin(\theta_0) + i_q\cos(\theta_0) \right] \sum_{s=0}^{S-1} \sin\left(\frac{2\pi s}{S}\right)\cos\left(\frac{2\pi s}{S}\right) \\[2ex]
-\left[ i_d\cos(\theta_0) - i_q\sin(\theta_0) \right] \sum_{s=0}^{S-1} \sin\left(\frac{2\pi s}{S}\right)\cos\left(\frac{2\pi s}{S}\right) + \left[ i_d\sin(\theta_0) + i_q\cos(\theta_0) \right] \sum_{s=0}^{S-1} \sin^2\left(\frac{2\pi s}{S}\right)
\end{bmatrix}
\\
\end{aligned}
$}
\end{equation*}

\noindent The discrete orthogonality properties of sinusoids over a full period of $S$-samples states that, due to linear independence, the summation of squared sinusoids evaluate to $\frac{S}{2}$ and that, due to linear dependence, the summation of the cross-products evaluates to $0$:
\begin{equation*}
\scalebox{0.55}{$\displaystyle
\sum_{s=0}^{S-1} \cos^2\left(\frac{2\pi s}{S}\right) = \frac{S}{2}, \quad
\sum_{s=0}^{S-1} \sin^2\left(\frac{2\pi s}{S}\right) = \frac{S}{2}, \quad
\sum_{s=0}^{S-1} \cos\left(\frac{2\pi s}{S}\right)\sin\left(\frac{2\pi s}{S}\right) = 0
$}
\end{equation*}

\noindent Applying these discrete orthogonality properties of sinusoids to the summation terms as previously rearranged produces:
\begin{equation*}
\scalebox{0.6}{$\displaystyle
\begin{aligned}
X_{i_a}(k=1) &= \frac{2}{S} \begin{bmatrix}
\left[ i_d\cos(\theta_0) - i_q\sin(\theta_0) \right] \cdot \frac{S}{2} - \left[ i_d\sin(\theta_0) + i_q\cos(\theta_0) \right] \cdot 0 \\[2ex]
-\left[ i_d\cos(\theta_0) - i_q\sin(\theta_0) \right] \cdot 0 + \left[ i_d\sin(\theta_0) + i_q\cos(\theta_0) \right] \cdot \frac{S}{2}
\end{bmatrix}
\\
&=\frac{2}{S} \begin{bmatrix}
\left[ i_d\cos(\theta_0) - i_q\sin(\theta_0) \right] \cdot \frac{S}{2} \\[2ex]
\left[ i_d\sin(\theta_0) + i_q\cos(\theta_0) \right] \cdot \frac{S}{2}
\end{bmatrix}
\end{aligned}
$}
\end{equation*}

\noindent Canceling out the scaling factors $\frac{2}{S}$ and $\frac{S}{2}$ concludes with the single harmonic component as:
\begin{equation}
\scalebox{0.6}{$\displaystyle
\begin{aligned}
& X_{i_a}(k=1) = 
\begin{bmatrix}
i_d\cos(\theta_0) - i_q\sin(\theta_0) \\[2ex]
i_d\sin(\theta_0) + i_q\cos(\theta_0)
\end{bmatrix}
\end{aligned}
$}
\label{eq:X1_Coeff}
\end{equation}

Finally, construct the frequency-domain representation. Ignoring the DC component as it is zero \eqref{eq:X0_Coeff}, we re-write the matrix formulation of $X_{i_a}$ \eqref{eq:X1_Coeff} as a complex Fourier coefficient at fundamental frequency, $\omega_{s0}$, to produce:
\begin{equation*}
\scalebox{0.85}{$\displaystyle
\begin{aligned}
& \tilde{I}_{\text{phA}} = 
[i_d\cos(\theta_0) - i_q\sin(\theta_0)] + j \cdot [i_d\sin(\theta_0) + i_q\cos(\theta_0)]
\end{aligned}
$}
\end{equation*}

\noindent Rearranging terms and applying complex algebra gives:
\begin{equation*}
\scalebox{0.75}{$\displaystyle
\begin{aligned}
\tilde{I}_{\text{phA}} &= i_d (\cos\theta_0 + j\sin\theta_0) - i_q \sin\theta_0 + j i_q \cos\theta_0 \\
&= i_d (\cos\theta_0 + j\sin\theta_0) + j^2 i_q \sin\theta_0 + j i_q \cos\theta_0 \\
&= i_d (\cos\theta_0 + j\sin\theta_0) + j i_q (\cos\theta_0 + j\sin\theta_0) \\
&= (i_d + j i_q)(\cos\theta_0 + j\sin\theta_0)
\end{aligned}
$}
\end{equation*}

\noindent Applying Euler's identity ($e^{j\theta_0} = \cos\theta_0 + j\sin\theta_0$) concludes with the phase-A current as:
\begin{equation}
\scalebox{1}{$\displaystyle
\begin{aligned}
\tilde{I}_{\text{phA}} &= (i_d + j i_q)e^{j\theta_0}
\end{aligned}
$}
\label{eq:euler_solution}
\end{equation}

\noindent Evidently, the frequency-domain grid quantity $\tilde{I}_{\text{phA}}$ is equal to the constant time-domain $dq$-quantities $i_{d/q}$ with a phase shift $\theta_0$. As formerly noted, other phases can be proved identically by first converting them to phase-A using \eqref{eq:SS_FD}. The final equations are summarized below:
\begin{subequations}
\label{eq:final_proof}
\begin{align}
    \tilde{I}_{\text{phA}} &= (i_d + j i_q) e^{j\theta_0} \label{eq:final_proof_a} \\
    \tilde{I}_{\text{phB}} &= (i_d + j i_q) e^{j\left(\theta_0 - \frac{2\pi}{3}\right)} \label{eq:final_proof_b} \\
    \tilde{I}_{\text{phC}} &= (i_d + j i_q) e^{j\left(\theta_0 + \frac{2\pi}{3}\right)} \label{eq:final_proof_c}
\end{align}
\end{subequations}
\qed

What has emerged is a general EMT frequency-domain model, constructed through three algebraic manipulations --- DC-constraining the $dq$-inputs, collapsing dynamic elements into memoryless components, and interfacing with the network via phase shift. These models, completely characterized by a memoryless $dq$-circuit \eqref{eq:step_2_conclusions} connected to the frequency-domain transmission grid via a phase shift \eqref{eq:final_proof}, are analyzed in the next Section~5 to demonstrate their computational efficiency and accuracy.

\section{Results}

Across device- to system-level evaluations, SALT matches EMT steady-state solution fidelity \cite{ParaEMT1,ParaEMT2} with a speedup of $1\times10^6$ times, whereas PF methods fail to capture security violations and introduce initialization errors. Performance is compared with traditional PF \cite{PF_constraints} and an enhanced formulation (enh-PF) \cite{PFfreqDev1} that incorporates distributed slack and static ZIP loads for abnormal operating conditions.

Because the steady-state power system problem can admit multiple solutions \cite{EqCircuit1}, a disparity between these tools could reflect convergence to different operating points rather than modeling errors. Thus, although omitted for brevity, we cross-substitute each tool's solution into the others' equations in each experiment, verifying the residuals are above solver tolerance and hence indicate true modeling errors rather than a different solution point.

We take SG parameters from ParaEMT \cite{ParaEMT1,ParaEMT2}, IBR parameters from an NREL benchmark \cite{GF_IBRs}, and load compositions and IM parameters from industrial loading \cite{loadComposition, Kundur}. EMT simulations use timesteps on the order of 10us.

We compare EMT steady-state results to SALT and PF using \eqref{unifyDef}, where $\mathcal{F}$ is evaluated as the one-sided DFT for tractability due its finite window and discrete nature. In lieu of $t$ at infinity, the DFT is computed when $v_t$ is periodic for two periods within a relative error margin of $0.1\%$.

\subsection{Disparities in Device Modeling}

We characterize how each tool models individual grid devices --- SGs, IBRs, IMs --- by comparing each of their steady-state responses against that of EMT device models with an alternating current (AC) source at the terminal, sweeping with varying magnitudes and frequencies. This establishes a device-level fidelity that underlies all subsequent system-level analyses.

EMT and SALT generator devices include a SG \cite{Kundur} and grid-forming (GFM) IBR \cite{GF_IBRs}. Figure~\ref{fig:sweepPV} shows the SALT error is flat across the AC current sweep, indicating its device response is consistent with EMT physics. Conversely, the constant PV generator used in PF yields a real power error with respect to frequency, attributed to its failure to model the droop mechanism in governor and PV-controls \cite{TGOV1,GF_IBRs}. The enh-PF generation model captures droop through distributed slack, adjusting the real power with respect to frequency, but it still incurs errors from missing non-idealities --- namely, resistive losses from the SG armature and GFM-IBR filter, or the shunt field rheostat setting behavior in the exciter \cite{excRheostate}. As a result, enh-PF experiences errors when linearized estimates do not reflect device nonlinearities.

EMT and SALT loads are composite models, comprising of a static ZIP and dynamic IM \cite{loadSummary}. Figure~\ref{fig:sweepPQ} indicates the SALT error is flat across an AC voltage source sweep of the IM. In contrast, the constant PQ load in PF fails to capture the voltage dependency of power from inductive coupling \cite{IM}. The ZIP load in enh-PF, fitted with quadratic voltage variation and linear frequency deviation, provides better alignment, but deviations persist by missing higher-order nonlinearities due to air gap power losses \cite{Kundur}. 

\par\addvspace{\intextsep}%
{\centering\captionsetup{type=figure}%
    \centering
    \begin{subfigure}{0.48\columnwidth}
        \centering
        \includegraphics[width=\textwidth]{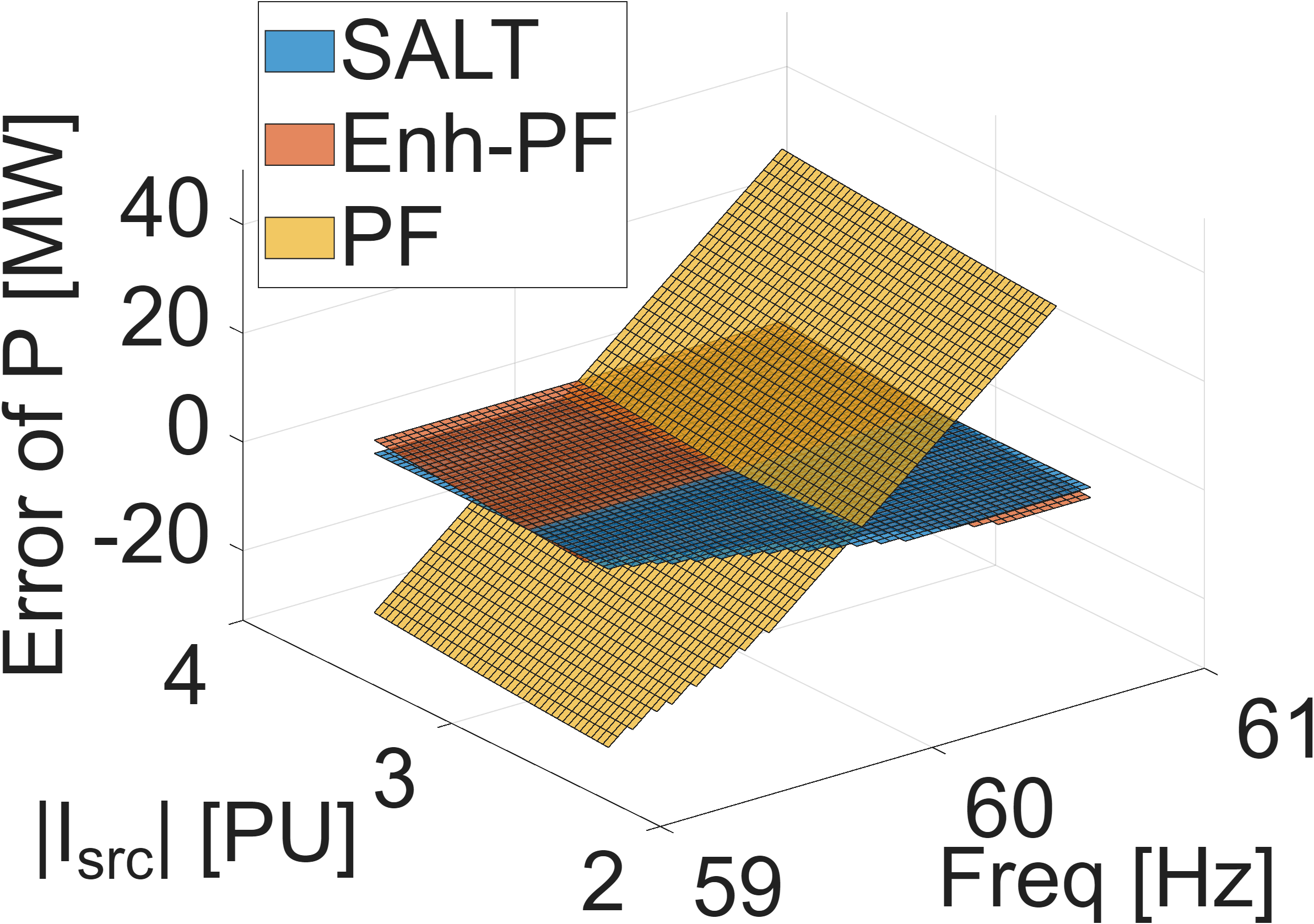}
        \caption{SG real power}
    \end{subfigure}
    \hfill
    \begin{subfigure}{0.48\columnwidth}
        \centering
        \includegraphics[width=\textwidth]{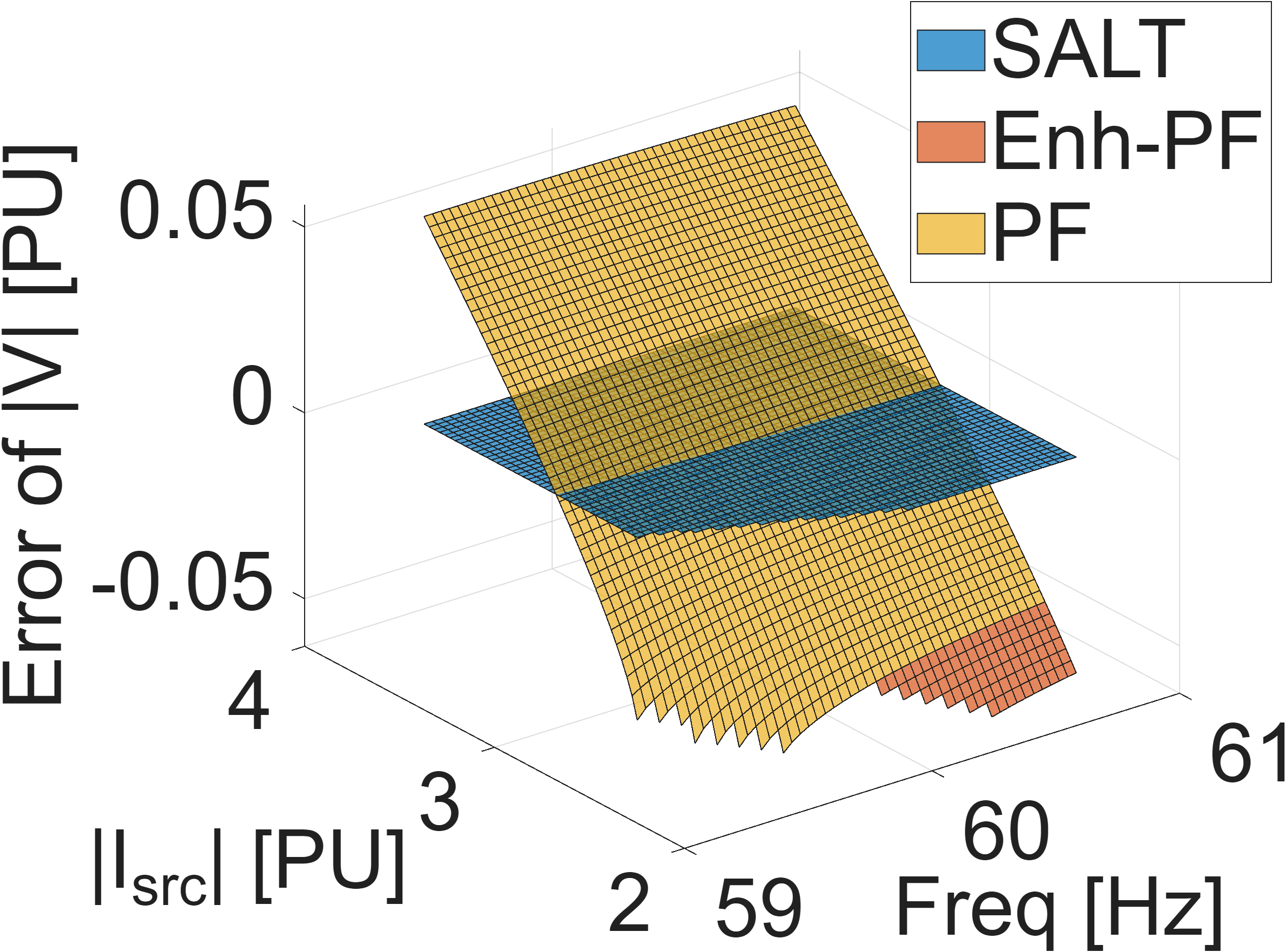}
        \caption{SG voltage}
    \end{subfigure}
    \centering
    \begin{subfigure}{0.48\columnwidth}
        \centering
        \includegraphics[width=\textwidth]{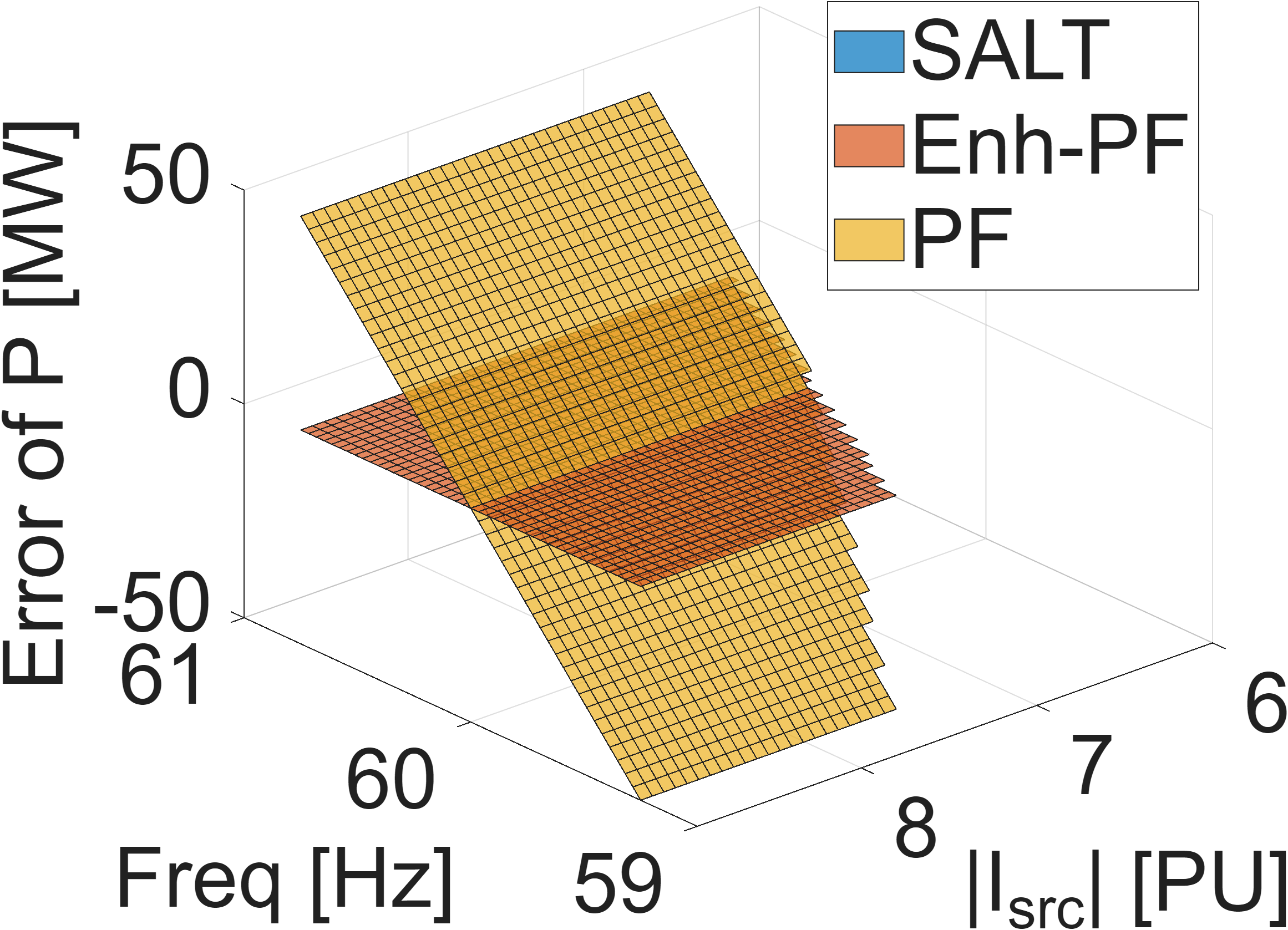}
        \caption{GFM-IBR real power}
    \end{subfigure}
    \hfill
    \begin{subfigure}{0.48\columnwidth}
        \centering
        \includegraphics[width=\textwidth]{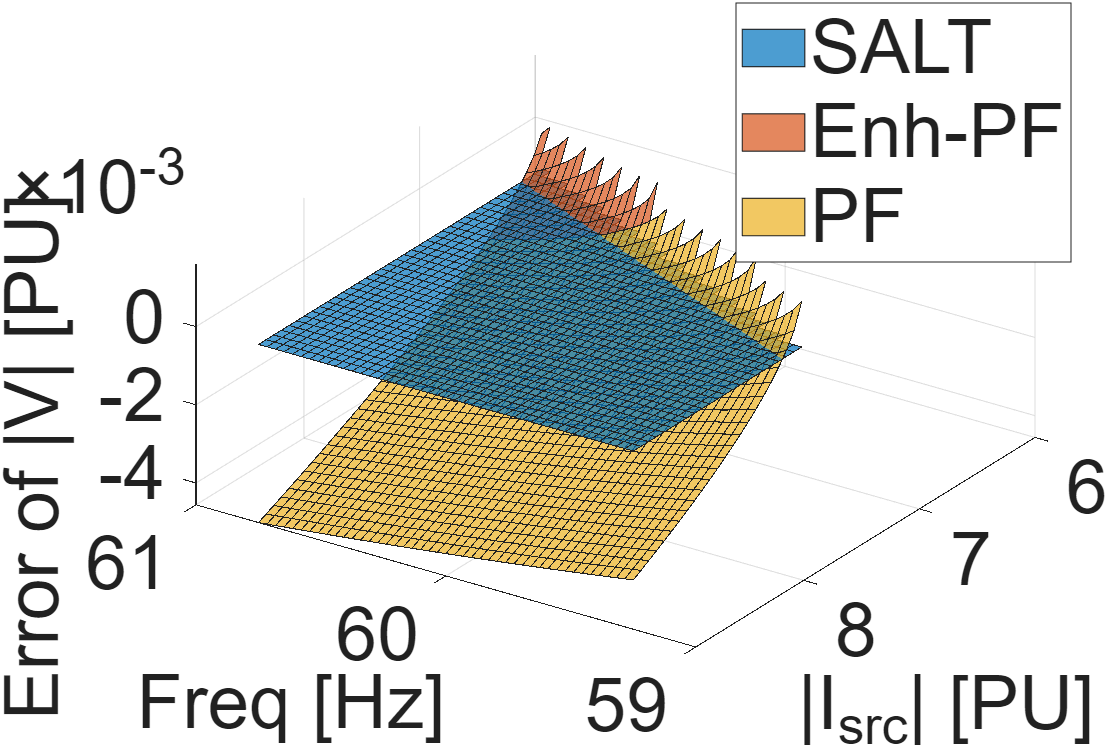}
        \caption{GFM-IBR voltage}
    \end{subfigure}
    \par\caption{\textit{\textbf{SALT generator models have error within numerical noise relative to EMT.} SALT (blue) error is flat. PF (yellow) frequency-dependent error is due to lack of droop control. Enh-PF (orange) incorporates droop through distributed slack but omits loss.}}
    \label{fig:sweepPV}
\par}
\addvspace{\intextsep}\par

\par\addvspace{\intextsep}%
{\centering\captionsetup{type=figure}%
    \centering
    \begin{subfigure}{0.48\columnwidth}
        \centering
        \includegraphics[width=\textwidth]{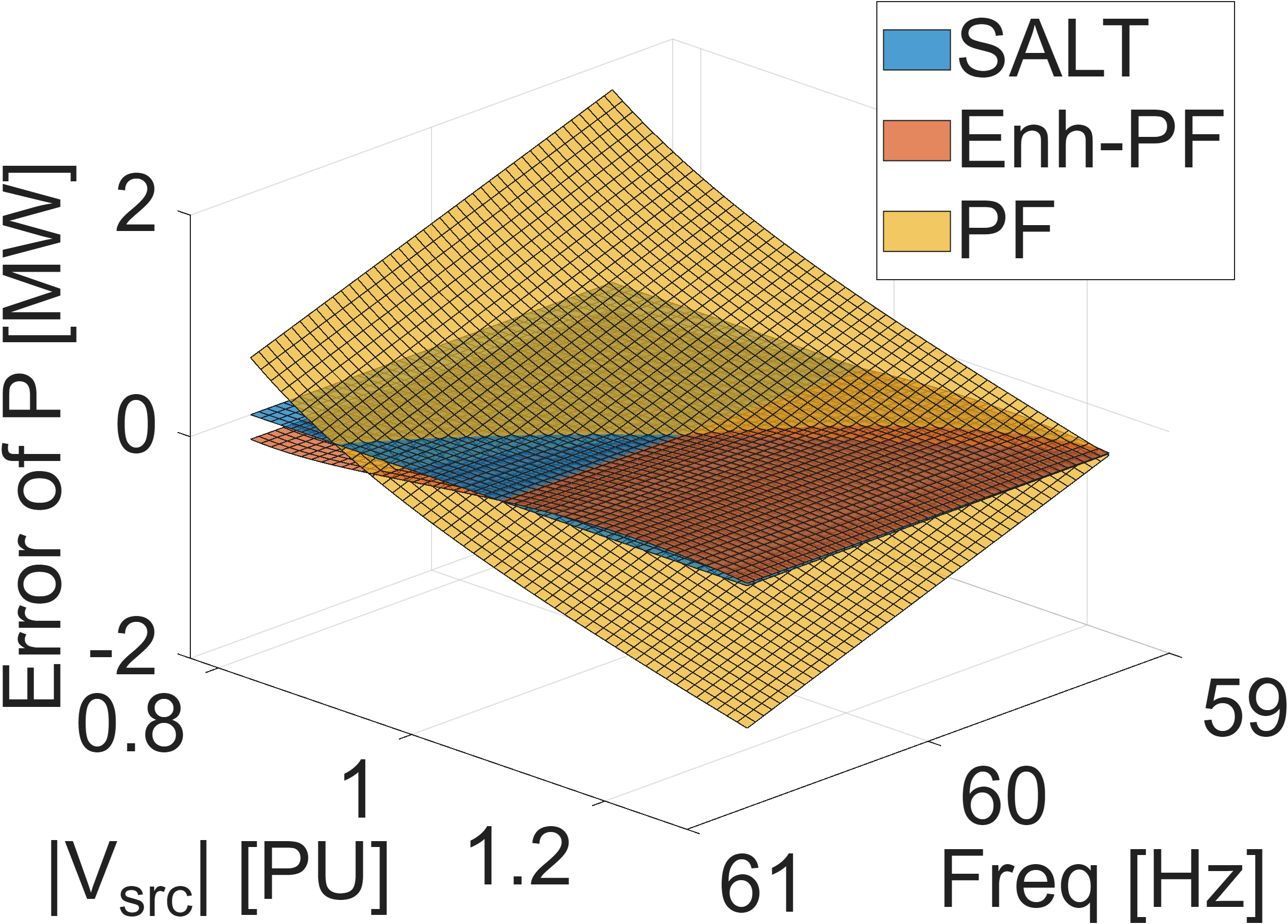}
        \caption{IM real power}
        \label{fig:first}
    \end{subfigure}
    \hfill
    \begin{subfigure}{0.48\columnwidth}
        \centering
        \includegraphics[width=\textwidth]{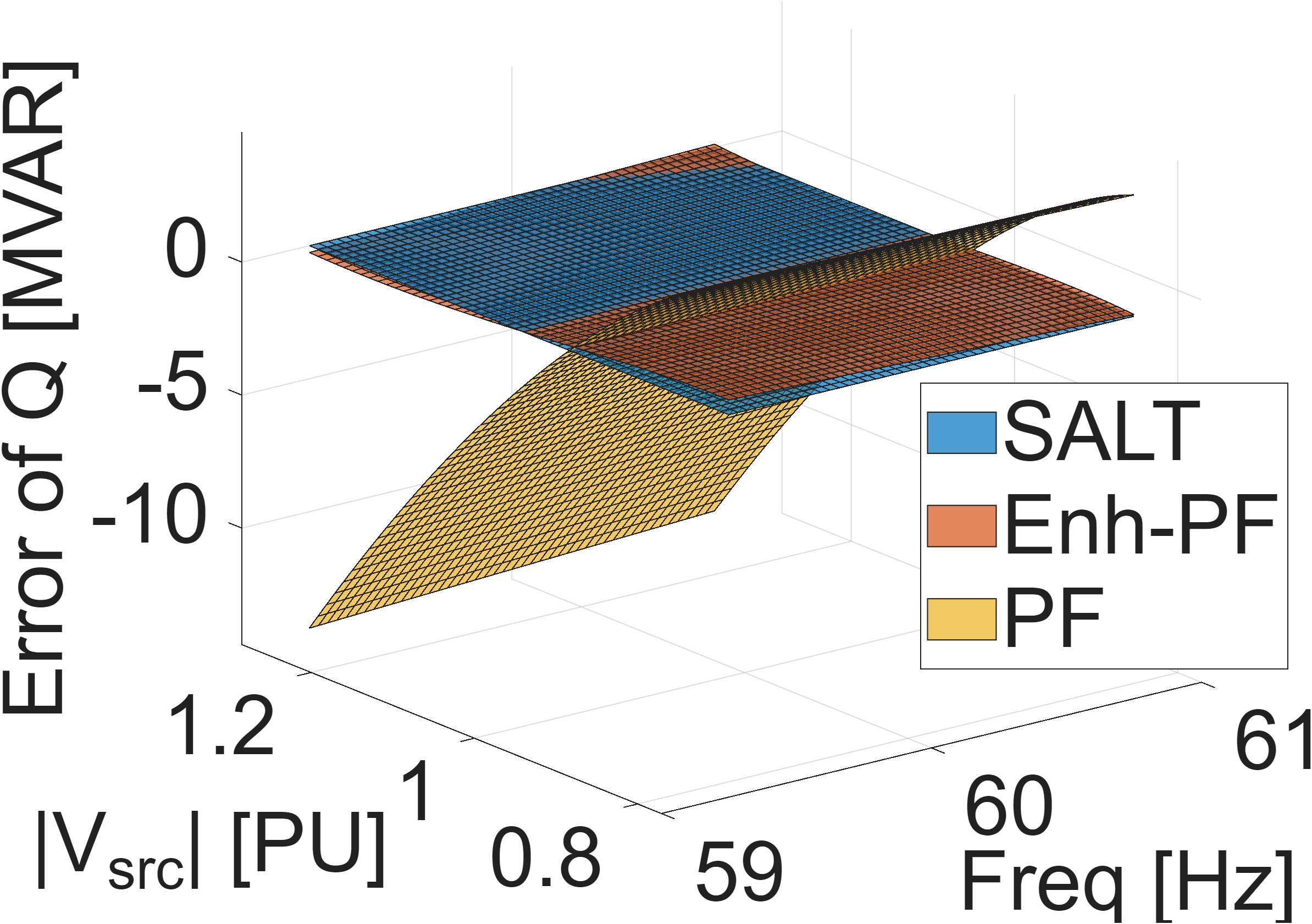}
        \caption{IM reactive power}
        \label{fig:second}
    \end{subfigure}
    \par\caption{\textit{\textbf{SALT IM in composite load model is consistent with EMT behavior.} SALT (blue) error is flat. PF (yellow) poorly models reactive power voltage dependency. Quadratic voltage fitting and linear frequency fitting in enh-PF (orange) fails to capture air gap power transfers and loss.}}
    \label{fig:sweepPQ}
\par}
\addvspace{\intextsep}\par


\subsection{Accuracy of System-Level Solutions}

With device-level error profiles established, we examine whether model errors remain localized or accumulate when the devices are embedded into transmission systems of increasing size. Figure~\ref{fig:largeSys} shows SALT and EMT solutions matching within numerical precision, with their deviations on the order of $1\times10^{-5}$ PU for voltage magnitude, $1\times10^{-3}$ degrees for voltage phase, and $1\times10^{-5}$ Hz for frequency. In contrast, PF errors grow with system size and do not average out, reaching a up to $0.187$ PU, $129$ degrees, and $0.786$ Hz. Enh-PF results in a smaller accumulation of errors of around $0.01$ PU, $0.4$ degrees, and $0.0014$ Hz. Despite not scaling with system size, these errors are consequential in obscuring violations, as demonstrated in the following Section~5.3.

\par\addvspace{\intextsep}%
{\centering\captionsetup{type=figure}%
    \centering
    \begin{subfigure}{\columnwidth}
        \centering
        \includegraphics[width=0.8\textwidth]{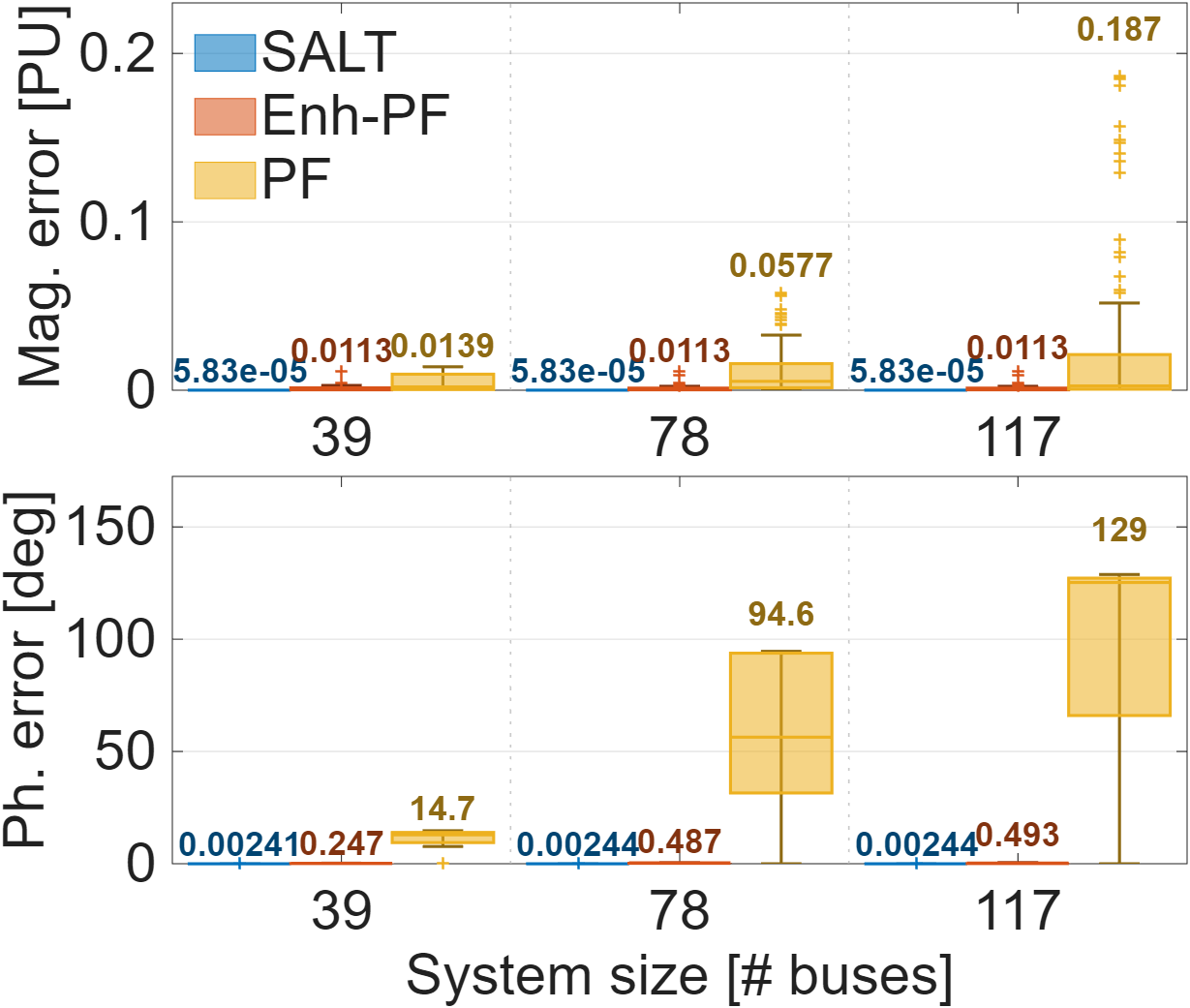}
        \caption{Bus voltage}
        \label{fig:figure1}
    \end{subfigure}
    \begin{subfigure}{\columnwidth}
        \centering
        \hspace{0.5cm}\includegraphics[width=0.8\textwidth]{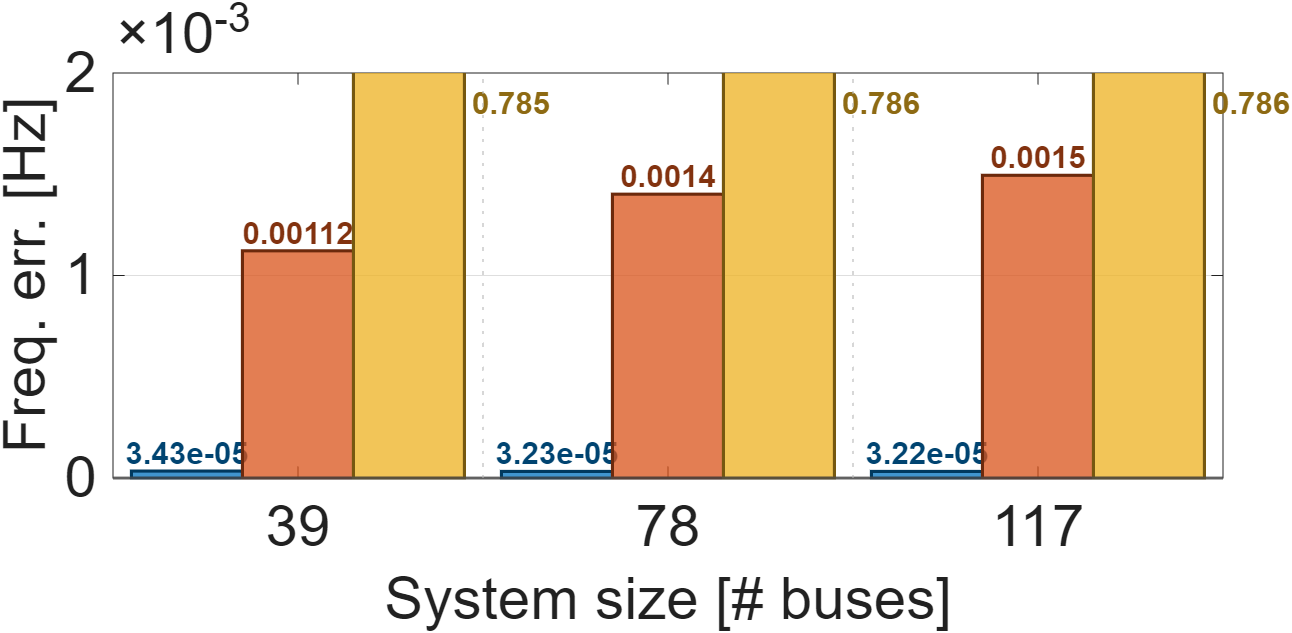}
        \caption{Frequency deviation}
        \label{fig:figure2}
    \end{subfigure}
    \par\caption{\textit{\textbf{EMT and SALT solutions have errors within numerical noise, whereas PF errors compound with system size.} Enlarged systems are formed by concatenating chains of the 39-bus network \cite{ParaEMT1,ParaEMT2} (all generating units are IBRs except at bus 30), stressed by the loss of generator at bus 34 and a 2\% load increase.}}
    \label{fig:largeSys}
\par}
\addvspace{\intextsep}\par

\subsection{Impact on Security Assessment}

The practical consequence of system-level errors is their impact on grid security assessment. Errors that compound across the network --- particularly under stressed, post-contingency conditions --- risk mischaracterizing real violations. To demonstrate this, an N-1 analysis for the loss of a generator is performed on the 39-bus system that was analyzed in the previous Section~5.2. The total violations computed by each tool is summarized in Table~\ref{tab:lossGen}.

The violations computed by SALT fully agree exactly with EMT simulation. Meanwhile, enh-PF understates the stress on the system, capturing 3 fewer voltage violations, 1 fewer Q-limit violation, and 1 fewer rated-line violation. PF underestimates the severity of rated-line overload, reporting 6 fewer violations, but mostly overstates the stress of the grid, computing a higher total of voltage and Q-limit violations and failing to converge to a feasible solution in the loss of the largest generator.

\par\addvspace{\intextsep}%
\noindent\begin{minipage}{\columnwidth}
    \centering
    \captionsetup{type=table}%
    \caption[]{\centering Contingency Analysis for 39-bus System with High Penetration of IBRs}
    \label{tab:lossGen}    
    \renewcommand{\arraystretch}{1.3} 
    \begin{tabular}{lcccc}
    \hline
    \raisebox{-4pt}{\textbf{Violation type}}
    & \multicolumn{3}{c}{\textbf{Number of violations}} \\ \cline{2-5}
     & \textbf{EMT} & \textbf{SALT} & \textbf{Enh-PF} & \textbf{PF} \\
    \hline
    \textbf{Q-limit} & 11 & 11 & 10 & 9 \\
    \textbf{Voltage} & 43 & 43 & 40 & 68 \\
    \textbf{Rated line} & 8 & 8 & 7 & 2 \\
    \textbf{No convergence} & 0 & 0 & 0 & 1 \\
    \hline
    \end{tabular}
    \begin{flushleft}
    \footnotesize *The number of violations in the table is the raw count. PF should be compared against the convergence-matched values i.e. number of EMT and SALT violations at only those contingencies where PF converged, which gives 5 Q-limit, 4 rated-line, 24 voltage violations. This reverses the Q-limit implication: PF actually computes more violations than EMT, not fewer, reinforcing its more stressed report of the grid.
    \end{flushleft}
    \vspace{0.1cm}
\end{minipage}
\par
\addvspace{\intextsep}\par

In this contingency analysis, the fewer rated-line violations reported by PF is not a straightforward result of PF device models producing less power. Figure~\ref{fig:tradPFvsSALT}a depicts negative \textit{and positive} errors in line power, indicating the accumulation of fewer violations are not simply attributed to PF computing \textit{less} power flowing through the network but rather a mischaracterized \textit{route} flow. The largest mischaracterization is concentrated within a specific set of lines, illustrated by the positive skew of errors in Figure~\ref{fig:tradPFvsSALT}a. Figure~\ref{fig:figureCA_reroutedPfPower} reveals that these specific lines of heavy flow are localized around the slack, which ties to how PF artificially resolves system power mismatches: by assuming a single compensator supplying infinite power, PF relieves line loading elsewhere, leaving rated-line violations in those regions under-reported. 

\par\addvspace{\intextsep}%
{\centering\captionsetup{type=figure}%
    \centering
    
    \begin{subfigure}[b]{0.495\columnwidth}
        \centering
        \includegraphics[width=\linewidth]{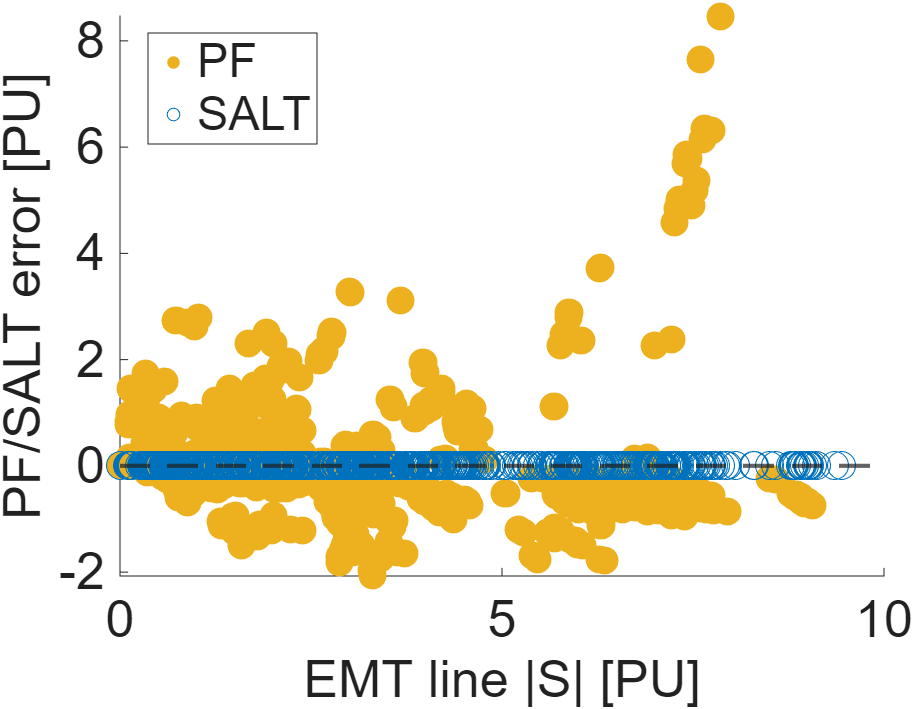}
        \caption{Line absolute power}
    \end{subfigure}
    \hfill 
    \begin{subfigure}[b]{0.495\columnwidth}
        \centering
        \includegraphics[width=\linewidth]{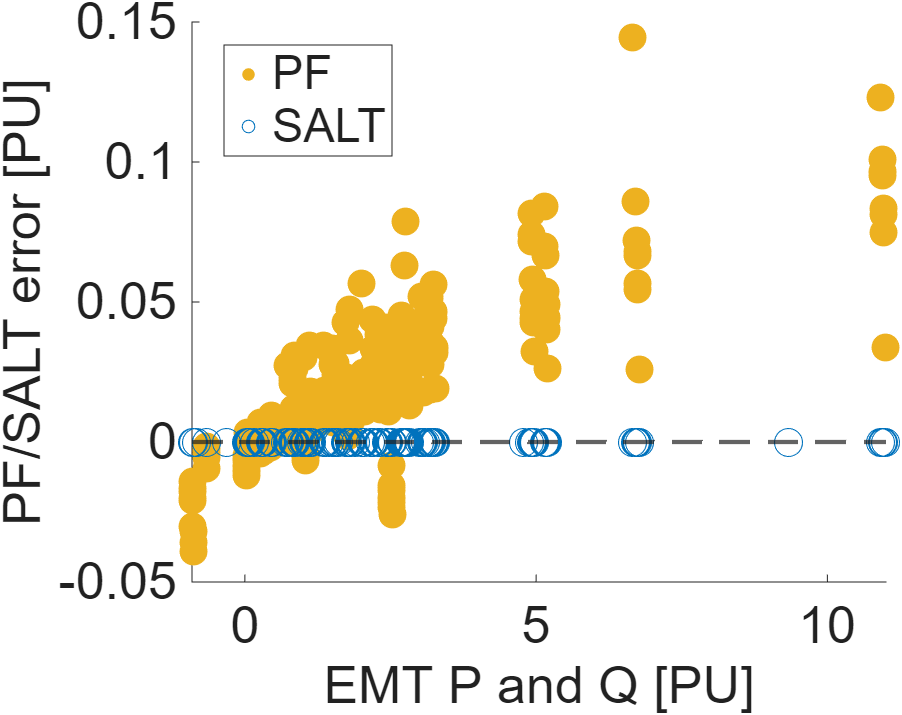}
        \caption{Load real and reactive power}
    \end{subfigure}
    \par\caption{\textit{\textbf{PF mischaracterizes the flow of power, artificially reducing and increasing loading on different lines (subfigure-a). PF also computes a more severe power deficit due to higher load consumption (subfigure-b).} This contingency analysis for the loss of generation is performed on the 39-bus system detailed in Figure~\ref{fig:largeSys}.}}
    \label{fig:tradPFvsSALT}
    
\par}
\addvspace{\intextsep}\par

\par\addvspace{\intextsep}%
{\centering\captionsetup{type=figure}%
\centering
\includegraphics[width=0.9\linewidth]{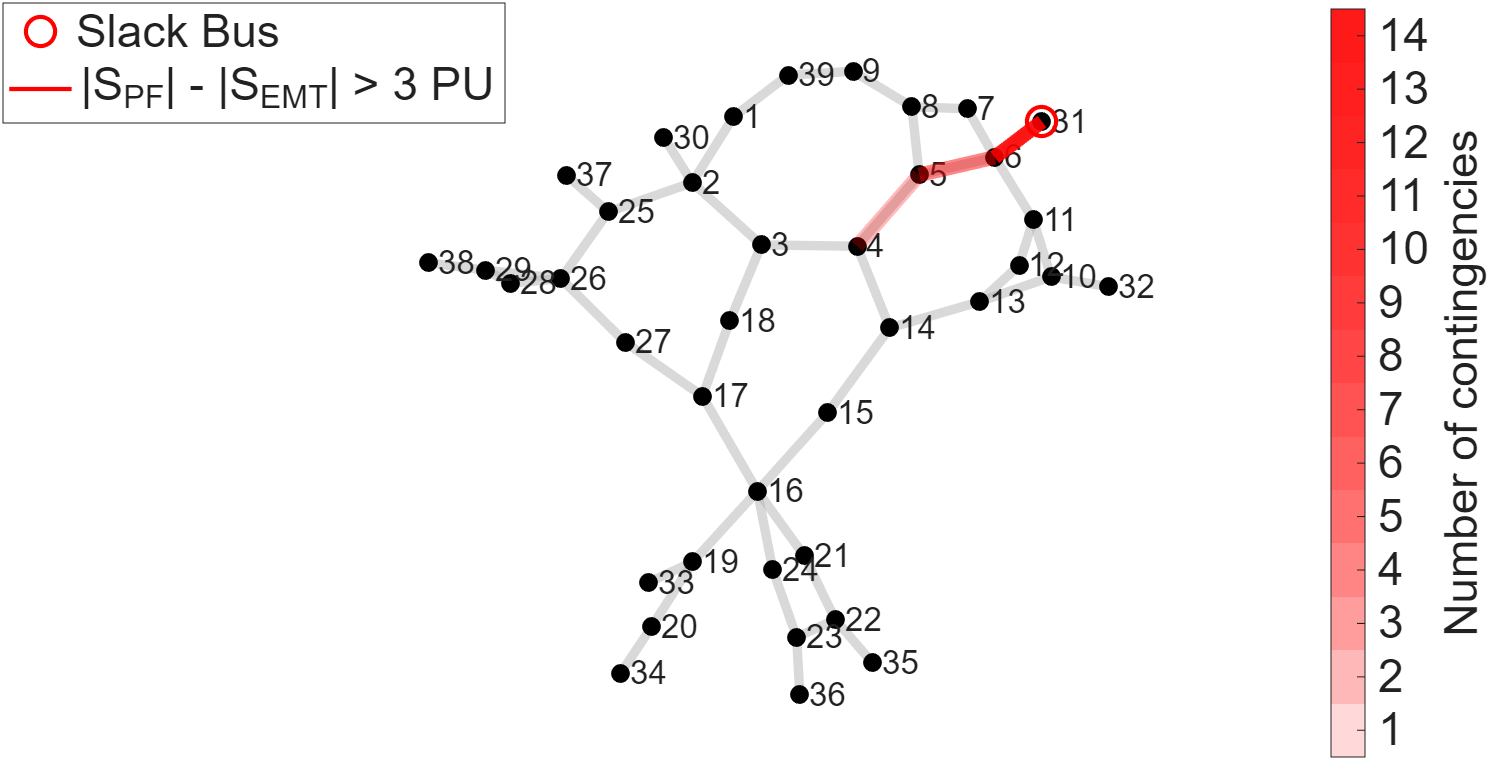}
\par\caption{\textit{\textbf{Large flows in PF are localized to the slack bus, consistent with the assumption that all power mismatches are compensated at a single node.} The heavy line load threshold 3 PU is selected based on the error magnitude of the skew in Figure~\ref{fig:tradPFvsSALT}a.}}
\label{fig:figureCA_reroutedPfPower}
\par}
\addvspace{\intextsep}\par

Additionally, PF is reporting a more stressed system, computing more voltage and Q-limit violations, as well as failing to converge to a feasible solution during the loss of the largest generator. The increased stress originates from the error of PF load models consuming more power, as shown by the positive P and Q errors in Figure~\ref{fig:tradPFvsSALT}b. In grids experiencing contingencies, IMs draw less power as a natural response to voltage dropping because the magnetic coupling weakens, whereas PF's stiff PQ bus remains at nominal power consumption (refer to positive error at lower voltage voltage in Figure~\ref{fig:sweepPQ}).

\subsection{Errors in EMT Initialization}

Modeling discrepancies between PF and EMT also degrade initialization for EMT simulation. Using the PF solution as the initial state injects errors that require transients to decay before simulation can begin. As shown in Table~\ref{tab:initBurden} on systems of increasing size stressed by a $25\%$ load increase, SALT initialization requires no transient correction, whereas PF-based initialization requires at least $100,000$ timesteps to settle, a burden that scales with network size. Although enh-PF computes solutions that are orders-of-magnitude closer to the EMT steady-state than PF (see previous Figures~\ref{fig:largeSys}, \ref{fig:4busFreq}), its number of settling timesteps do not decrease for EMT initialization. Figure~\ref{fig:4busFreq} explains why: despite a closer initial frequency, its first swing reaches the same magnitude as PF. As such, the smaller initialization error computed by enh-PF does not translate into a shorter transient simulation, underscoring the need for SALT as an exact steady-state solver.

\noindent\begin{minipage}{\columnwidth}
    \centering
    \captionsetup{type=table}%
    \caption{EMT Initialization}
    \label{tab:initBurden}
    \renewcommand{\arraystretch}{1.3} 
    \begin{tabular}{lcccc}
    \hline
    \raisebox{-4pt}{\textbf{   System   }}
    & \multicolumn{3}{c}{\textbf{Timesteps (10\textmu s) until periodic}} \\ \cline{2-4}
     & \textbf{SALT} & \textbf{Enh-PF} & \textbf{PF} \\
    \hline
        \textbf{4-bus} & 
        \begin{tabular}{@{}c@{}}0\end{tabular} &
        \begin{tabular}{@{}c@{}}136,680\end{tabular} &
        \begin{tabular}{@{}c@{}}135,005\end{tabular} \\
        \textbf{8-bus} &
        \begin{tabular}{@{}c@{}}0\end{tabular} &
        \begin{tabular}{@{}c@{}}141,705\end{tabular} &
        \begin{tabular}{@{}c@{}}146,060\end{tabular} \\
        \textbf{12-bus} &
        \begin{tabular}{@{}c@{}}0\end{tabular} &
        \begin{tabular}{@{}c@{}}141,370\end{tabular} &
        \begin{tabular}{@{}c@{}}156,780\end{tabular} \\
        \textbf{16-bus} &
        \begin{tabular}{@{}c@{}}0\end{tabular} &
        \begin{tabular}{@{}c@{}}141,705\end{tabular} &
        \begin{tabular}{@{}c@{}}161,470\end{tabular} \\
        \hline
    \end{tabular}
    \begin{flushleft}
    \footnotesize *Enlarged systems are formed by concatenating chains of the 4-bus network \cite{MATPOWER} (generation at bus 4 is IBR), stressed by $25\%$ load increase.
    \end{flushleft}
    \vspace{0.1cm}
\end{minipage}

\par\addvspace{\intextsep}%
{\centering\captionsetup{type=figure}%
    \vspace{-0.1cm} 
    \centering
    \begin{subfigure}[b]{0.75\columnwidth}
        \centering
        \includegraphics[width=\linewidth]{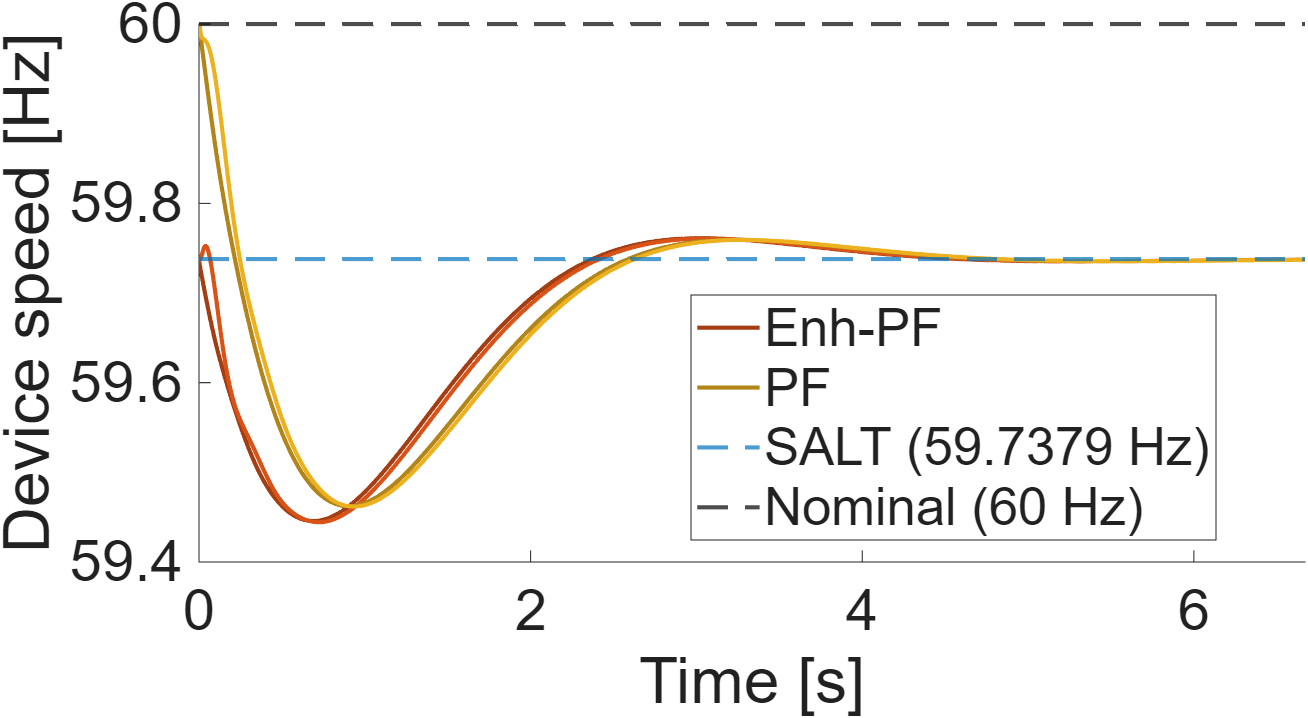}
    \end{subfigure}
    \par\caption{\textit{\textbf{PF and enh-PF initializations both require $\sim$6\,s to settle, despite enh-PF providing a closer starting value.} Initialization is evaluated on the 4-bus system \cite{ParaEMT1,ParaEMT2} under a 25\% load increase with GFM-IBR displacing generation at bus 4. The long settling time is consistent with load-step contingencies simulated on comparable low-inertia systems \cite{frequencySettle}.}}
    \label{fig:4busFreq}
\par}
\addvspace{\intextsep}\par

\subsection{Computation and Runtime}

The computation and runtime for simulating the perturbed systems in the previous Section~5.4 are summarized in Table~\ref{tab:computation}. Here, each tool is initialized with identical pre-stressed conditions. EMT runtimes are around $1\times10^6$ times slower than PF. The runtimes of PF-based methods and SALT are on the same order-of-magnitude, with SALT efficiently maintaining 3-4 NR iterations across expanded networks while PF grows to 6 iterations. PF requires more NR steps because its final solution is farther from the initial state (previously illustrated in Figure~\ref{fig:largeSys}).

\noindent\begin{minipage}{\columnwidth}
    \centering
    \captionsetup{type=table}%
    \caption{Runtime}
    \label{tab:computation}
    
    \renewcommand{\arraystretch}{1.3} 
    \setlength{\tabcolsep}{3.5pt} 

    \resizebox{\columnwidth}{!}{
    \begin{tabular}{llcccc}
        \hline
        \textbf{System} & 
        \textbf{Metric} & 
        \multicolumn{4}{c}{\textbf{Computation}} \\ \cline{3-6}
        & & \textbf{EMT} & \textbf{SALT} & \textbf{Enh-PF} & \textbf{PF} \\
        \hline
        
        \multirow{3}{*}{\textbf{4-bus}} 
        & Total runtime & 6,151,100 & 1.965 & 2.486 & 1 \\
        \cline{2-6}
        & Timesteps & 736,560 & - & - & - \\
        \cline{2-6}
        & NR iterations & 3 (avg.) & 3 & 4 & 4 \\
        \hline 
        
        \multirow{3}{*}{\textbf{8-bus}} 
        & Total runtime & 5,943,600 & 2.6777 & 1.789 & 1 \\
        \cline{2-6}
        & Timesteps & 731,538 & - & - & - \\
        \cline{2-6}
        & NR iterations & 3 (avg.) & 4 & 5 & 5 \\
        \hline
        
        \multirow{3}{*}{\textbf{12-bus}} 
        & Total runtime & 5,787,300 & 1.9747 & 1.2658 & 1 \\
        \cline{2-6}
        & Timesteps & 739,908 & - & - & - \\
        \cline{2-6}
        & NR iterations & 3 (avg.) & 4 & 5 & 6 \\
        \hline
        
        \multirow{3}{*}{\textbf{16-bus}} 
        & Total runtime & 5,867,200 & 2.059 & 1.2266 & 1 \\
        \cline{2-6}
        & Timesteps & 728,190 & - & - & - \\
        \cline{2-6}
        & NR iterations & 3 (avg.) & 4 & 6 & 6 \\
        \hline
    \end{tabular}%
    }
    \begin{flushleft}
    \footnotesize *Runtime is normalized as a ratio to that of PF.
    \end{flushleft}
    \vspace{0.1cm}
\end{minipage}

However, the total runtime of SALT is higher than PF, maximally reaching $2.7\times$ in Table~\ref{tab:computation}, which is attributed to the speed of matrix construction and solve in each NR evaluation. SALT evaluates more sophisticated nonlinear device equations, incorporates frequency-dependency into branch models (e.g. linear capacitor in PF becomes nonlinear in SALT), and solves a larger matrix expanded by device internal states. Because the cost of matrix factorization dominates in large problems for steady-state solvers \cite{McCalla}, the computation of SALT is confined to the complexity of LU factorization, $\mathcal{O}(n^3)$, and in practice, is smaller for sparse problems, such as the empirical $\mathcal{O}(n^{1.1}\text{--}n^{1.5})$ for circuits~\cite{Pillage}. Figure~\ref{fig:FINAL_LARGE_SYS_RUNTIME} shows that on the large benchmark cases~\cite{MATPOWER}, SALT reaches at most $7\times$ PF's compute time.

\par\addvspace{\intextsep}%
{\centering\captionsetup{type=figure}%
    \vspace{-0.1cm} 
    \centering
    \begin{subfigure}[b]{1.01\columnwidth}
        \centering
        \includegraphics[width=\linewidth]{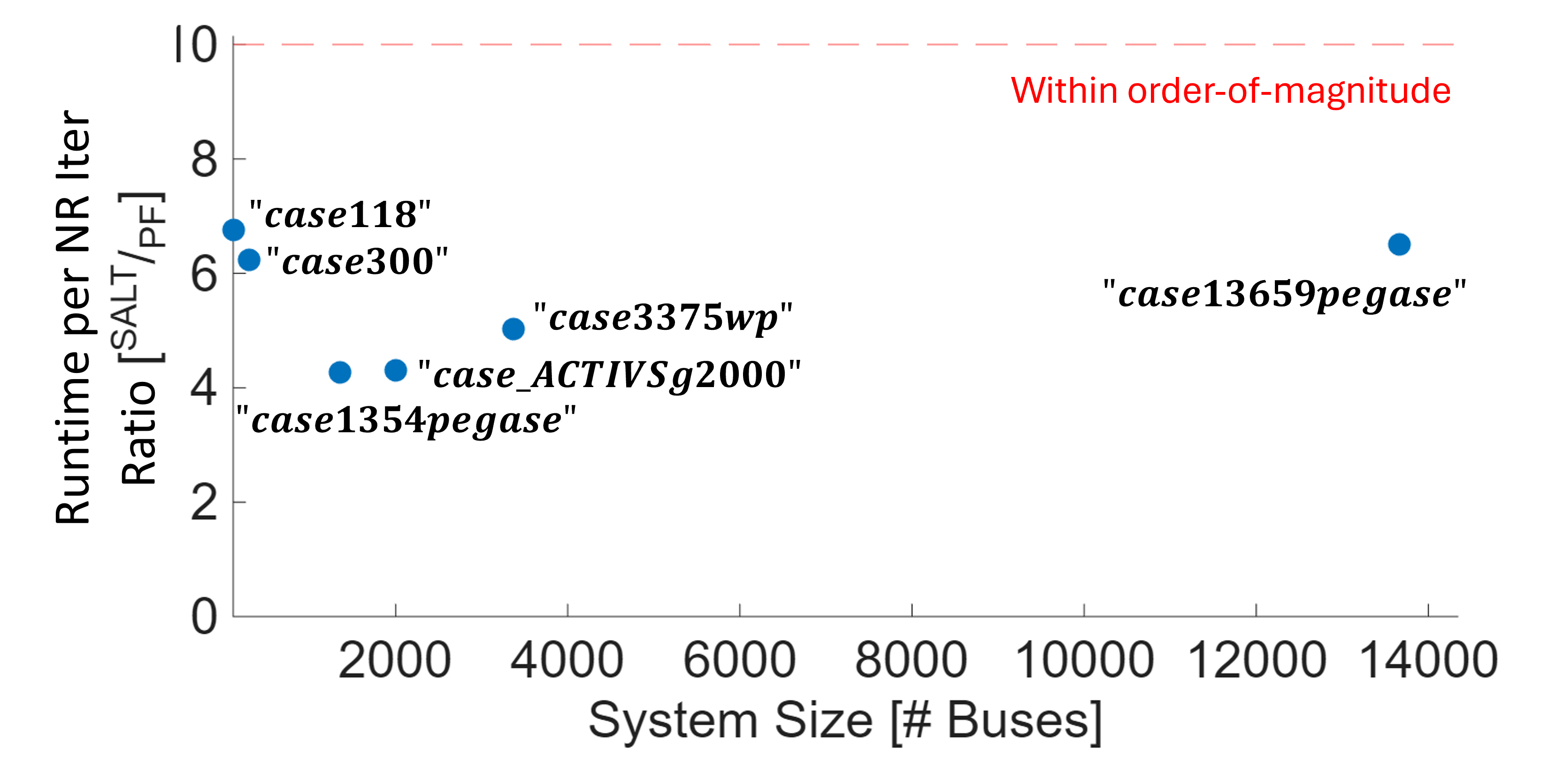}
    \end{subfigure}
    \par\caption{\textit{\textbf{The per iteration runtime of SALT remains order of magnitude to that of PF.} Unlike the total runtime used in Section 5.5, the runtime per NR iteration is the average runtime across the number of iterations to evaluate the speed of matrix construction and solve, not the number of steps to convergence.}}.
    \label{fig:FINAL_LARGE_SYS_RUNTIME}
\par}
\addvspace{\intextsep}\par

\section{Code Availability}
\begin{center}
\textit{https://github.com/tinagaostrawberry/SALT}
\end{center}

\section{Conclusion}
We propose unifying steady-state analysis and EMT simulation with SALT, a physics-based framework that directly solves EMT steady-state systems while maintaining the computational speed of PF. The methodology derives frequency-domain equivalents for dynamic nonlinear devices and embeds them into transmission network analysis.

The necessity of unified steady-state analysis is highlighted by the failure of PF models at scale. PF errors do not average out but compound with system size, yielding deviations up to 0.168 PU in our results. Our contingency analysis highlights the consequences of PF errors: while SALT captures post-outage violations consistently, PF-based solvers reported 75\% fewer rated line violations, 18\% fewer Q-limit violations, and 7\% fewer voltage violations. The burden of EMT simulation is costly, taking hundreds-of-thousands of timesteps and requiring $1\times10^6$ times more runtime. SALT achieves strict physical accuracy while maintaining runtimes on the same order-of-magnitude as PF.

\section{Declaration of Generative AI and AI-assisted Technologies in the Manuscript Preparation Process}

During the preparation of this work, the author(s) used AI (Anthropic, Opus 5. 2026) to generate the graphical abstract. After using this tool/service, the author(s) reviewed and edited the content as needed and take(s) full responsibility for the content of the published article.

\printcredits

\bibliographystyle{elsarticle-num}

\end{document}